\documentclass[%
 aip,
 amsmath,amssymb,
 reprint,%
]{revtex4-1}

\usepackage{graphicx}% Include figure files
\usepackage{dcolumn}% Align table columns on decimal point
\usepackage{bm}% bold math
\usepackage[utf8]{inputenc}
\usepackage[T1]{fontenc}
\usepackage{mathptmx}
\usepackage{etoolbox}
\usepackage{braket}
\usepackage{amsmath,amsthm}
\usepackage{amssymb} % for \mathbb
\usepackage[english]{babel}
\usepackage{latexsym} 
\usepackage{amsmath,amsthm}
\usepackage{ifpdf}
\usepackage{epstopdf}
\usepackage{subfig}
\usepackage{soul}
\usepackage{dcolumn}% Align table columns on decimal point
\usepackage{bm}% bold math
\usepackage{amssymb} % for \mathbb
\usepackage{braket}
\usepackage{wrapfig}
\usepackage[dvipsnames]{xcolor}
\usepackage{color}
\usepackage[colorlinks=true, allcolors=blue]{hyperref}
\usepackage{graphicx}

\newcommand{\beq}{\begin{equation}}
\newcommand{\eeq}{\end{equation}}
\newcommand{\barr}{\begin{eqnarray}}
\newcommand{\earr}{\end{eqnarray}}
\newcommand{\bseq}{\begin{subequations}}
\newcommand{\eseq}{\end{subequations}}
\newcommand{\expectation}[3]{\langle #1|#2|#3\rangle}

\newcommand{\vett}[1]{\textbf{#1}}

\usepackage[colorlinks=true, allcolors=blue]{hyperref}

\usepackage[normalem]{ulem}

\makeatletter
\def\@email#1#2{%
 \endgroup
 \patchcmd{\titleblock@produce}
  {\frontmatter@RRAPformat}
  {\frontmatter@RRAPformat{\produce@RRAP{*#1\href{mailto:#2}{#2}}}\frontmatter@RRAPformat}
  {}{}
}%
\makeatother

\begin{document}

\preprint{AIP/123-QED}

\title{Single-Photon Detection via Enhanced Cross-Phase Modulation in Topology-Optimized Epsilon-Near-Zero Dual-Wavelength Nanocavities}
% Force line breaks with \\
\author{Luca Dal Negro}
 \email{dalnegro@bu.edu.}
\affiliation{Department of Electrical \& Computer Engineering, Boston University, 8 Saint Mary's Street, Boston, 02215, MA, USA}
\affiliation{Department of Physics, Boston University, 590 Commonwealth Avenue, Boston,02215, MA, USA}
\affiliation{Division of Materials Science \&  Engineering, Boston University, 15 St. Mary’s street, Brookline, 02446, MA, USA}%Lines break automatically or can be forced with \\
\author{Riccardo Franchi}%
\affiliation{Department of Electrical \& Computer Engineering, Boston University, 8 Saint Mary's Street, Boston, 02215, MA, USA}

\author{Marco Ornigotti}
 \affiliation{Faculty of Engineering and Natural Sciences, Tampere University, Tampere, Finland} 

\date{\today}% It is always \today, today,

\begin{abstract}
We use the Green's tensor quantization theory for open resonant nanostructures with absorption losses to study cross-phase modulation (XPM) at the single-photon level in nanoscale Kerr-type epsilon-near-zero (ENZ) materials with an effective nonlinear susceptibility integrated inside dual-wavelength nanocavities.
We obtain analytical formulas for the XPM frequency shifts in hybrid nanocavities that simultaneously trap a classical probe beam at 1.5 $\mu$m and single-photon pump at 3 $\mu$m wavelengths.
We present a comprehensive analysis of the fundamental limits for mid-infrared single-photon detection in the quantum nondemolition modality for nanostructured cadmium oxide (CdO) regions with ENZ-enhanced nonlinearity embedded in a silicon (Si) environment inversely designed by free-form topology optimization.
We numerically implement our theoretical results using finite element simulations within the rigorous framework of quasi-normal modes, demonstrating a single-photon XPM frequency shift $\Delta f_s \approx 5.71 \text{ GHz}$ with fractional shift (i.e., frequency pulling) $\Delta f_s / f_s \approx 2.85 \times 10^{-5}$ and addressing the feasibility of detection in the hybrid Si-CdO dual-wavelength nanocavity, either with a classical probe beam or a squeezed probe state, including the contributions of traditional limitations from self-phase modulation noise, thermorefractive noise, shot noise, and free-carrier absorption effects.
Finally, we present a comparative size scaling analysis of the XPM phase shift and phase noise contributions for dual-wavelength nanocavities based on CdO and indium tin oxide (ITO) nonlinear ENZ materials.
This work establishes a robust benchmark for the engineering of mid-infrared single-photon nonlinear devices such as nondemolition quantum detectors, sensors, and all-optical gates on a solid state photonic platform.
\end{abstract}

\maketitle

\section{Introduction}
\noindent Single-photon nonlinear quantum optics using
solid-state nanostructures compatible with silicon (Si) technology bear the promise to revolutionize
quantum information technologies by providing scalable and energy-efficient solutions for the realization of optically controlled quantum gates, entangled photon sources, and integrated quantum nondemolition detection (QND) devices capable to detect individual photons across large frequency spectra at room-temperature (RT) \cite{munro2005high,ferretti2012single,carusotto2010feshbach,tonndorf2017chip,kirchmair,shapiro2006single,choi2017self,eisaman2011invited,dal2025nonlinear}. Presently, single-photon nonlinearities have been demonstrated using Rydberg atoms in ultrahigh-Q cavities at cryogenic temperatures, which suffer from limited bandwidth and scalability, besides significant challenges for  integration with photodetectors \cite{chang2014quantum}. Unfortunately, nonlinear optical effects at the single-photon level on a scalable solid-state platform cannot be achieved using traditional nonlinear materials or conventional photonic structures, motivating the development of novel approaches that leverage extreme light-matter coupling and nonlinear phenomena at the nanoscale. 
Recently, the inverse design of photonic devices based on free-form topology optimization (TopOpt) techniques has gained relevance for developing increasingly efficient photonic nanostructures with deep sub-wavelength mode confinements and large quality factors, resulting in dramatic enhancements of light-matter interactions at the nanoscale using scalable and Si-compatible materials \cite{christiansen2021inverse, kang_large-scale_2024, albrechtsen2022nanometer, ma_inverse-designed_2023, albrechtsen2022nanometer,jensen2011topology, ouyang2024singular, xiong2024experimental, dal2025nonlinear}.
Moreover, the discovery of enhanced nonlinear optical responses in epsilon-near-zero (ENZ) tunable materials, such as indium tin oxide (ITO) and high mobility cadmium oxide (CdO), provides additional opportunities for high density integration of active photonic components on a Si-based platform \cite{reshef2019nonlinear, reshef2017beyond, alam2016large, capretti2015enhanced, capretti2015comparative, shubitidze2024enhanced, shubitidze2025enhancement, schrecengost2024large, runnerstrom2017epsilon, yang2019high, dal2025nonlinear}. 

In this paper, we investigate mid-infrared RT detection of single photons in topology optimized dual-wavelength nanocavities using rigorous macroscopic quantum electrodynamics (QED) analysis in the framework of the Green's tensor quantization theory for open resonant structures with absorption losses.
In particular, motivated by the large and sub-picosecond fast nonlinear optical responses dominated by free-carrier relaxations in ITO and CdO ENZ materials \cite{KinseyReview}, we develop a general theoretical framework to study enhanced cross-phase modulation (XPM) at the single-photon level in ENZ nanostructures integrated in dual-wavelength nanocavities and obtain general analytical formulas for the achievable XPM frequency shift in optimized resonant nanostructures that simultaneously trap a classical probe (signal) beam at 1.5 $\mu$m and a single pump photon at 3 $\mu$m.
We also present a comprehensive noise analysis of the fundamental detection limits for high mobility cadmium oxide (CdO) and ITO nonlinear media with ENZ behaviour embedded in optimized Si or silicon nitride (SiN) environments and we address the feasibility of single-photon detection considering the realistic limitations that arise from self-phase modulation noise, thermorefractive noise, shot noise, and electronic jitter effects.
We numerically implement our theoretical results using finite element simulations within the rigorous framework of quasi-normal modes and demonstrate a single-photon XPM frequency shift $\Delta f_s \approx 5.71 \text{ GHz}$ in a CdO cavity region with diameter $D_{\rm ENZ}=20\,{\rm nm}$.  
Our findings establish a robust benchmark for the design and engineering of mid-infrared single-photon nonlinear devices on the Si platform, such as nondemolition quantum detectors, sensors, and all-optical gates on a solid state photonic platform.

Our paper is organized as follows: in Sec.~\ref{general} we briefly review the quantization approach for linear and Kerr-type nonlinear dispersive materials, respectively, which are described by an effective $\chi^{(3)}$ susceptibility, on which our subsequent analysis will be based.
In Sec.~\ref{XPMshift} we obtain closed-form general expressions for the XPM frequency shift in a general medium, while in  Sec.~\ref{nanocavity} we apply the theory to resonant nanostructures within the rigorous framework of quasi-normal modes.
In Sec.~\ref{sec:topopt}, we describe in detail the free-form optimization procedure to achieve dual-wavelength Si-CdO nanocavities and in Sec.~\ref{feasibility} we address the feasibility of nondemolition single-photon detection at RT via a comprehensive noise analysis that includes the contributions of the self-phase-modulation (SPM), the thermorefractive, and the free-carrier absorption.
In Sec.~\ref{sec:scaling} we discuss the size scaling analysis of XPM-driven single-photon detection in both CdO- and ITO-based nanocavities.
Finally, in Sec.~\ref{conclusions}, we draw our conclusions.

\section{General Green's function quantization theory in lossy media} \label{general}
\noindent The Green's function quantization approach, also known as macroscopic quantum electrodynamics, rigorously models quantum light in lossy, causal, and dispersive media. By avoiding specific microscopic models of matter, it rigorously expresses the quantized electromagnetic field in terms of the classical Green's tensor, inherently satisfying the fluctuation-dissipation theorem (FDT) \cite{vogel2006quantum}. To quantize the electromagnetic field in this general framework, the medium is treated not just as a background, but as a coupled system of light and matter (i.e., polaritonic excitations). The classical electric field \(\mathbf{E}(\mathbf{r}, \omega)\) inside the medium obeys the macroscopic wave equation driven by a microscopic polarization noise current $\mathbf{j}_N(\mathbf{r}, \omega)$ according to:
\begin{equation}
	\nabla \times \nabla \times \mathbf{E}(\mathbf{r}, \omega) - \frac{\omega^2}{c^2}\, \epsilon(\mathbf{r}, \omega)\, \mathbf{E}(\mathbf{r}, \omega) = i \omega\, \mu_0\, \mathbf{j}_N(\mathbf{r}, \omega).
\end{equation}

In a lossy, dispersive material, the electric field operator \(\hat{\mathbf{E}}(\mathbf{r}, \omega)\) cannot be expanded into simple plane-wave modes. Instead, in the Green's function quantization approach we must use the noise polarization operator \(\hat{\mathbf{f}}(\mathbf{r}, \omega)\) and the classical dyadic Green's function \(\mathbf{G}(\mathbf{r}, \mathbf{r}', \omega)\) defined as \cite{vogel2006quantum}:
\begin{equation}\label{eq1}
\nabla\times\nabla\times\overleftrightarrow{\mathbf{G}}(\mathbf{r},\mathbf{r}^{\prime},\omega)-\frac{\omega^2}{c^2}\varepsilon(\mathbf{r},\omega)\overleftrightarrow{\mathbf{G}}(\mathbf{r},\mathbf{r}^{\prime},\omega)={\delta}(\mathbf{r}-\mathbf{r}^{\prime})\overleftrightarrow{\mathbb{I}},
\end{equation}
where $\overleftrightarrow{\mathbb{I}}$ denotes the unit dyadic.
The quantized electric field operator can be then written as
\begin{equation}\label{eq2}
\hat{\mathbf{E}}(\mathbf{r},t)=\int_0^{\infty}\,d\omega\,\left[\hat{\mathbf{E}}^+(\mathbf{r},\omega) + \hat{\mathbf{E}}^-(\mathbf{r},\omega)\right],
\end{equation}
where
\begin{equation}\label{eq3}
\hat{\mathbf{E}}^{+}(\mathbf{r},\omega )=i\int d^{3}\mathbf{r}'\,c(\vett{r}',\omega)\,\overleftrightarrow{\mathbf{G}}(\mathbf{r},\mathbf{r}^{\prime },\omega )\cdot \hat{\mathbf{f}}(\mathbf{r}^{\prime },\omega ),
\end{equation}
with $c(\vett{r}',\omega)=\omega^2\sqrt{\hbar\operatorname{Im}[\varepsilon(\vett{r}',\omega)]}/\left(c^2\sqrt{\pi\varepsilon_0}\right)$, and
$\hat{\mathbf{E}}^-(\mathbf{r},\omega)=\left[\hat{\mathbf{E}}^+(\mathbf{r},\omega)\right]^{\dagger}$. The polariton noise operators $\hat{\mathbf{f}}(\mathbf{r},\omega)$ are continuous-mode bosonic operators satisfying the commutation relations
\begin{equation}
	\left[\hat{\mathbf{f}}_{i}(\mathbf{r},\omega ),\hat{\mathbf{f}}_{j}^{\dag }(\mathbf{r}^{\prime },\omega ^{\prime })\right]=\delta _{ij}\delta (\mathbf{r}-\mathbf{r}^{\prime })\delta (\omega -\omega ^{\prime }),
\end{equation}
 and represent the absorbing degrees of freedom of the medium (i.e., the heat bath). Remarkably, in this framework the FDT requires that the imaginary part of the Green's function (the dissipative part) is directly proportional to the vacuum quantum fluctuations of the field via \cite{scheel2006causal}
\begin{equation}
\langle\hat{E}_{j}(\mathbf{r},\omega )\hat{E}_{k}^{\dagger}(\mathbf{r}^{\prime },\omega ^{\prime })\rangle =\frac{\hbar \omega ^{2}}{\pi \epsilon _{0}c^{2}}\text{Im}\,G_{jk}(\mathbf{r},\mathbf{r}^{\prime },\omega )\delta (\omega -\omega ^{\prime }).
\end{equation}

Notice that in the Green's function approach causality is inherently built-in through the analytical properties of the Green's tensor, and through the fact, that $\epsilon(\mathbf{r},\omega)$ satisfies the Kramers-Kronig relations. Moreover, the specific boundary conditions of the electromagnetic environment are directly accounted by $\overleftrightarrow{\mathbf{G}}(\mathbf{r},\mathbf{r}',\omega)$, which naturally encapsulates all open boundary conditions. This feature is particularly important in modeling the dissipative nature of open cavities, as it rigorously takes into account radiative leakage and out-coupling to the far-field, treating the cavity and the surrounding environment as a unified quantum system. Finally, by linking material absorption (\(\text{Im}\,\epsilon(\mathbf{r},\omega)\)) directly to the continuous polariton noise operators (\(\hat{\mathbf{f}}\)) via the FDT, the Green's function quantization framework guarantees that the fundamental quantum commutation relations, such as the Heisenberg uncertainty principle, are preserved at every point in space and time, regardless of the amount of cavity losses.

\subsection{Slowly Varying Fields and Coherent States}

\noindent In this work, we consider slowly-varying fields, i.e., fields that can be written as
\begin{equation}
\hat{\textbf{E}}^+(\textbf{r},t)=e^{-i\omega_0t}\,\hat{\textbf{E}}^+_{\text{SVAA}}(\textbf{r},t;\omega_0),
\end{equation}
where $\omega_0$ is the carrier (central) frequency. The slowly-varying amplitude approximation (SVAA) can be applied to Eq. \eqref{eq2} by introducing the SVAA polaritonic operator
\begin{equation}
\hat{\mathbf{h}}(\mathbf{r},t;\omega_0)=\frac{1}{\sqrt{\Delta\omega_0}}\int_{\Delta\omega_0}d\omega\,\hat{\mathbf{f}}(\mathbf{r},\omega)\,e^{i\omega_0t},
\end{equation}
where $\Delta\omega_0$ is the spectral bandwidth around $\omega_{0}$ (with $\Delta\omega_0\ll\omega_0$). The SVAA field operator then reads
\begin{equation}\label{eq8}
\hat{\mathbf{E}}^+_{\text{SVAA}}(\mathbf{r},t;\omega_0)=i\int d^3r'\,d(\vett{r}',\omega_0)\,\overleftrightarrow{\mathbf{G}}(\mathbf{r},\mathbf{r}',\omega_0)\cdot\hat{\mathbf{h}}(\mathbf{r}';\omega_0),
\end{equation}
where $d(\vett{r}',\omega_0)=\sqrt{\Delta\omega_0}\,c(\vett{r}',\omega_0)$. In the remaining of this work, for the sake of simplicity, we will drop the subscript SVAA from all field operators, always implicitly assuming the validity of this approximation.

To properly describe an electromagnetic coherent state, moreover, we introduce the polaritonic coherent states $\ket{\alpha(\mathbf{r}';\omega)}$ and, in analogy with standard quantum optics \cite{loudon2000quantum}, we assume they are the eigenstates of the polaritonic SVAA annihilation operator, i.e.,
\begin{equation}
	\hat{h}_{i}(\mathbf{r}^{\prime};\omega)\ket{\alpha(\mathbf{r}^{\prime};\omega)}=h_{i}(\mathbf{r}^{\prime};\omega)\ket{\alpha(\mathbf{r}^{\prime};\omega)},
\end{equation}
from which we can define the $i$-th component of the classical field as:
\barr
	E_{i}(\mathbf{r},t)&\equiv&\bra{\alpha}\hat{E}^+_{i}(\mathbf{r},t;\omega_0)\ket{\alpha}\nonumber\\
    &=&i\int\,d^3r'\,d(\vett{r}',\omega_0)\mathbf{G}_{ij}(\mathbf{r},\mathbf{r}',\omega_0)h_{j}(\mathbf{r}',\omega_0),
\earr
and, analogously, $E^*_i(\vett{r},t)=\expectation{\alpha}{\hat{E}_i^-(\vett{r},t;\omega_0)}{\alpha}$. In complete analogy to the lossless case, we posit that $E_{i}(\mathbf{r},t)=|\alpha|u_{i}(\mathbf{r},t)$, where $u_{i}(\mathbf{r},t)$ is a classical electric field shape function that must be chosen according to the specific problem at hand, and $|\alpha|^2$ is the average number of photons contained in the field \cite{gerry2023introductory}.

\section{Frequency shift in cross-phase modulation processes} \label{XPMshift}
In this section, we derive the XPM frequency shift due to a single-photon pump $\ket{1}_p$ and a classical coherent probe (signal) beam $\ket{\alpha}_s$.
The XPM is modelled by treating one field (the pump) as a source that modifies the refractive index experienced by a second field (the probe), all while accounting for the local density of states (LDOS) and losses provided by the Green's function. Specifically, XPM is driven by the causal third-order nonlinear polarization \cite{wilhelmi}
\begin{widetext}
\begin{equation}
\hat{\mathbf{P}}_{\\text{NL}}^+(\mathbf{r}, t)=\epsilon_0\int\,d\tau_1d\tau_2d\tau_3\,\boldsymbol\chi^{(3)}(\mathbf{r},t-\tau_1,t-\tau_2,t-\tau_3):\hat{\mathbf{E}}^+(\mathbf{r},\tau_1)\,\hat{\mathbf{E}}^-(\mathbf{r},\tau_2)\,\hat{\mathbf{E}}^+(\mathbf{r},\tau_3),
\end{equation}
\end{widetext}
generated by the interaction Hamiltonian \cite{boyd_nonlinear_2020,vogel2006quantum}
\begin{equation}\label{eq13}
\hat{H}_{\text{int}}=\frac{1}{4}\int _{V}d^{3}\mathbf{r}\,\hat{\mathbf{E}}^-(\mathbf{r},t)\cdot\hat{\mathbf{P}}_{\\text{NL}}^+(\mathbf{r},t)+ \text{H.c.}
\end{equation}
If we now substitute the Ansatz
\begin{equation}
\hat{\mathbf{E}}^+(\mathbf{r},t)=\hat{\mathbf{E}}^+_p(\mathbf{r},t\,;\omega_p)\,e^{-i\omega_p t}+\hat{\mathbf{E}}^+_s(\mathbf{r},t\,;\omega_s)\,e^{-i\omega_s t},
\end{equation}
into Eq. \eqref{eq13}, where subscripts $(p,s)$ indicate the pump and signal (probe) fields, respectively, and isolate the XPM term, we obtain
\barr
\hat{P}_{\text{XPM},i}^+(\mathbf{r},\omega _{s})&=&6\,\epsilon _{0}\sum _{j,k,l}\chi _{ijkl}^{(3)}(\mathbf{r},-\omega _{s};\,\omega _{s},-\omega _{p},\omega _{p})\,\hat{E}_{s,j}^+(\mathbf{r},\omega _{s})\nonumber\\
&\times&\hat{E}_{p,k}^-(\mathbf{r},\omega _{p})\hat{E}_{p,l}^+(\mathbf{r},\omega _{p}),
\earr
for the nonlinear polarization, and
\barr
\hat{\mathcal{H}}_{\text{XPM}}(\mathbf{r})&=&6\,\epsilon _{0}\sum_{i,j,k,l}\,\chi _{ijkl}^{(3)}(\mathbf{r},-\omega _{s};\,\omega _{s},-\omega _{p},\omega _{p})\,\hat{E}_{s,i}^-(\mathbf{r},\omega _{s})\nonumber\\
&\times&\hat{E}_{s,j}^+(\mathbf{r},\omega _{s})\,\hat{E}_{p,k}^-(\mathbf{r},\omega _{p})\,\hat{E}_{p,l}^+(\mathbf{r},\omega _{p}),
\earr
for the XPM interaction Hamiltonian density. Notice, that in deriving these expressions we did not choose any particular operator ordering, but left the ``physical ordering" $E_s^{\dagger}E_sE_p^{\dagger}E_p$, reminiscent of the fact that XPM depends on the intensity of signal/probe and pump fields. 

The XPM frequency shift can then be directly derived from the interaction Hamiltonian in a way similar to the derivation of the Kerr frequency shift per photon \cite{ourPRB2025,haroche,kirchmair}, using the Hellmann-Feynman theorem applied to a quantized field \cite{feynman1939forces}.
In our context, the XPM frequency shift is obtained as the derivative of the interaction Hamiltonian with respect to the probe energy $E_s = N_s \hbar \omega_s$ as follows \cite{louisell1972quantum,imoto1985quantum}:
\begin{equation}
	\Delta \omega _{1,ph}=\frac{1}{\hbar }\frac{\partial \langle \hat{H}_{\text{int}}\rangle }{\partial N_s},
\end{equation}
where $N_s$ is the average number of photons in the probe/signal classical state and $\langle \hat{H}_{\text{XPM}}\rangle=\bra{\psi}\hat{H}_{\text{XPM}}\ket{\psi}$
with $\ket{\psi}=\ket{1}_{p}\otimes\ket{\alpha}_{s}$. We also note that the total interaction Hamiltonian is
\begin{equation}
	\hat{H}_{\text{int}}=\int d^{3}r \hat{\mathcal{H}}_{\text{XPM}}(\mathbf{r}).
\end{equation}

Using the results above for the expectation values of the signal and pump field operators along with the commutation properties of the boson noise operators we obtain, after some straightforward algebra, the expectation value $\expectation{\Psi}{\hat{H}_{\text{int}}}{\Psi}$ of the interaction Hamiltonian
\begin{equation}\label{eq23}
\langle\hat{H}_{\text{int}}\rangle=\int\,d^3r\,\chi_{kl}^{(eff)}(\vett{r},\omega_p,\omega_s)E^{\ast}_{s,k}(\mathbf{r},\omega_{s})E_{s,l}(\mathbf{r},\omega_{s}).
\end{equation}
where
\barr
\chi^{(eff)}_{kl}(\vett{r},\omega_p,\omega_s)&=&\frac{3\hbar\omega_p^2\Delta\omega_p}{2\pi c^{2}}\chi _{ijkl}^{(3)}(\mathbf{r};-\omega _{s},\omega _{s},-\omega _{p},\omega _{p})\nonumber\\
&\times&\operatorname{Im}{G}_{ij}(\mathbf{r},\mathbf{r},\omega_p),
\earr
is the effective (i.e., vacuum-dressed) nonlinear susceptibility). We note how, in the general case of an anisotropic medium, the shift depends on the Directional (or Projected) LDOS. The components \(\text{Im}\,G_{ij}(\vett{r},\vett{r},\omega_p)\)  dictate that the pump field only alters the probe's frequency if their respective directional vacuum or material field fluctuations couple directly through a non-vanishing element of the \(\chi _{ijkl}^{(3)}\) tensor.
Expressing the probe/signal field as $|\vett{E}_{s}(\vett{r},t)|^{2}=|\alpha|^{2}|\vett{u}_{s}(\vett{r},t)|^{2}\equiv N_s|\vett{u}_{s}(\vett{r},t)|^{2}$, and taking the derivative with respect to the number $N_s$ of probe photons, we obtain the XPM frequency shift induced by the single-photon pump as
\barr\label{cross}
	\Delta \omega _{\text{XPM}}&=&\frac{3\,\omega_p^2\Delta\omega_p}{2\pi c^2}\int d^{3}r\,	\chi^{(3)}(\mathbf{r};\,\omega_{p};\,\omega_{s})\nonumber\\
    &\times&\text{Tr}\left[\text{Im}\,\mathbf{G}(\mathbf{r},\mathbf{r},\omega _p)\right]|\vett{u}_{s}(\vett{r},\omega_{s})|^{2},
\earr

where the contractions over the spatial indices collapsed into the trace (Tr) of the  dyadic tensor because here we assumed an isotropic material for simplicity. The trace of the imaginary part of the Green's function at identical spatial coordinates, \(\text{Tr}[\text{Im}\,\mathbf{G}(\mathbf{r}, \mathbf{r}, \omega)]\), is directly proportional to the electromagnetic Local Density of States (LDOS) that quantifies vacuum fluctuations. Therefore, the XPM frequency shift is ultimately a localized cross-modulation of the pump LDOS interacting with the probe field via the local \(\chi ^{(3)}\) nonlinearity in the structured medium.
\begin{figure}[!t]
    \centering
    \includegraphics[width=0.95\linewidth]{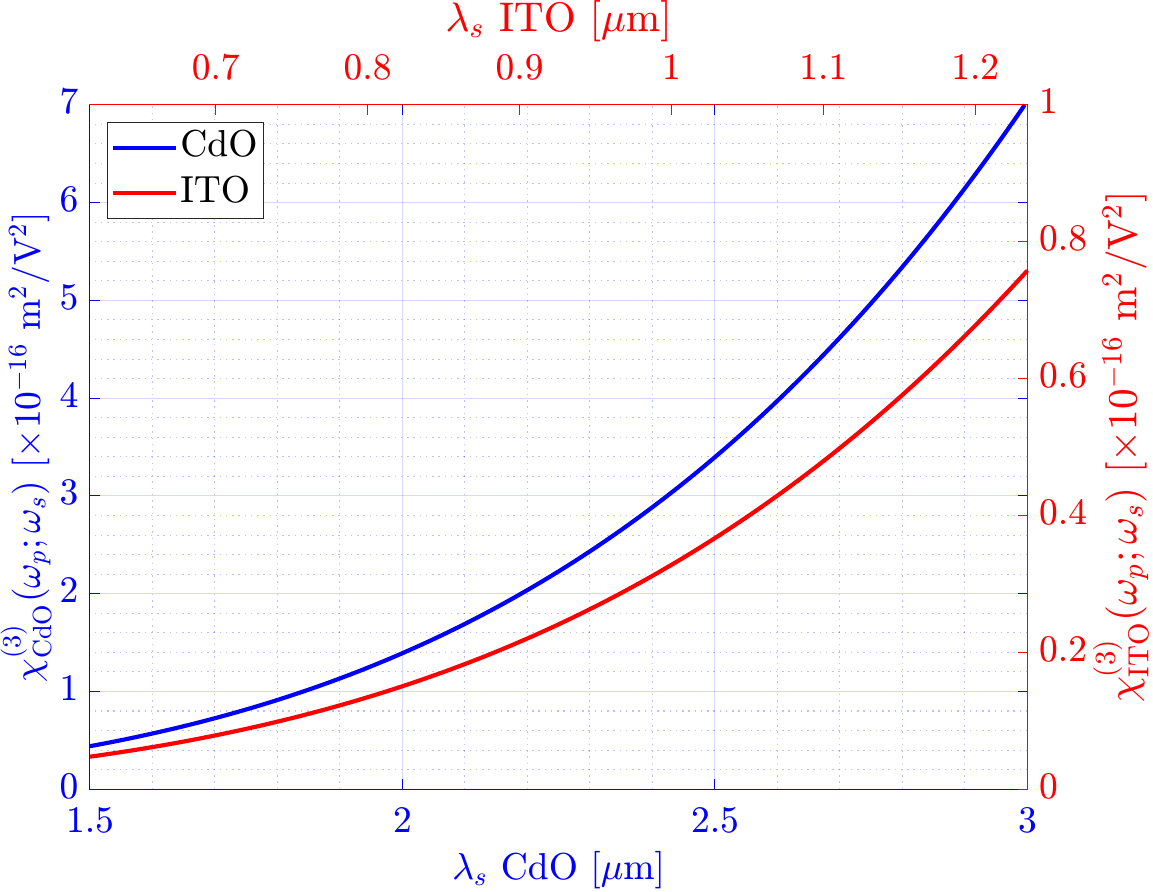}
    \caption{
    Third-order nonlinear susceptibility $\chi^{(3)}(\mathbf{r};\omega_{p};\omega_{s})$ as a function of $\lambda_s$ for CdO and ITO. The parameters used for the calculations are $\chi^{(3)}(\mathbf{r};\omega_{p};\omega_{p})=7\times 10^{-16}$ and $A = -(8.6970 + i\,1.5636)\times10^{-19}$ for CdO, and $\chi^{(3)}(\mathbf{r};\omega_{p};\omega_{p})=7.58\times 10^{-17}$ and $A = -(3.0966 + i\,1.6825)\times10^{-19}$ for ITO.}
    \label{fig:chi3_p_s}
\end{figure}
Following Ref. \cite{kashef2026path}, we can represent the general nonlinear susceptibility in its factorized form, reminiscent of the Miller's rule
\begin{widetext}
\begin{equation}\label{eq:Miller}
	\chi^{(3)}(\mathbf{r};\omega_{4};\omega_{3};\omega_{2};\omega_{1})=A\,\chi^{(1)}(\vett{r};\omega_{4})\,\chi^{(1)}(\vett{r};\omega_{3})\,\chi^{(1)}(\vett{r};\omega_{2})\,\chi^{(1)}(\vett{r};\omega_{1}),
\end{equation}
\end{widetext}
where $A$ is a normalization constant with the dimensions of $m^2/V^2$, and $\omega_{4}=\omega_{1}+\omega_{2}+\omega_{3}$.
For the XPM process at hand, we set $\omega_{1}=\omega_{p}$ and $\omega_{2}=\omega_{s}$ (which fixes $\omega_{3}=-\omega_{p}$ and $\omega_{4}=\omega_{s}$) and obtain
\begin{equation}
	\chi^{(3)}(\mathbf{r};\omega_{p};\omega_{s})=A[\chi^{(1)}(\vett{r};\omega_{s})]^{2}
	\chi^{(1)}(\vett{r};\omega_{p})\chi^{(1)}(\vett{r};-\omega_{p}),
\end{equation}
and, we fix $A$ in order to recover the measured value \cite{gulino2003large} $\chi^{(3)}(\mathbf{r};\omega_{p};\omega_{p})=7\times 10^{-16}$ $\text{m}^{2}/\text{V}^{2}$ at the pump (ENZ) frequency of the material $\omega_{p}=\omega_{\rm ENZ}=\sqrt{\omega^{2}_{0}/\epsilon_{b}-\gamma^{2}}$.
Figure~\ref{fig:chi3_p_s} shows the $\chi^{(3)}(\mathbf{r};\omega_{p};\omega_{s})$ as a function of the probe wavelength for both CdO and ITO ENZ materials that will be used in our nanocavity simulations.
We note that the strength of the cross phase modulation decreases significantly for both materials at shorter wavelengths. In particular, for CdO it reduces by a factor $\approx 15.9$ at $\lambda_s=1.5\,\mu\text{m}$ respect to its ENZ value at $\lambda_p=3.0\,\mu\text{m}$ and similarly for ITO where it drops by a factor of $15.94$ at $\lambda_s=0.617\,\mu\text{m}$ from its ENZ wavelength $\lambda_p=1.234\,\mu\text{m}$.
It is also useful to introduce the fractional shift, or frequency pulling $\delta$ of the cavity, as the ratio of the XPM frequency shift to the resonant frequency of the cavity signal
\barr
\delta\equiv\frac{\Delta\omega_{\text{XPM}}}{\omega_s}&=&\frac{3\omega_p^2\Delta\omega_p}{2\pi\omega_s c^2}\int\,d^3r\,	\chi^{(3)}(\mathbf{r};\omega_{p};\omega_{s})\nonumber\\
&\times&\operatorname{Tr}\operatorname{Im}\mathbf{G}(\mathbf{r},\vett{r},\omega_p)|\mathbf{u}_{s}(\mathbf{r},\omega_{s})|^{2}.
\earr
\section{Frequency shift in dual-band nanocavities at a single-photon level} \label{nanocavity}
We now apply our theory to the case of a sub-wavelength nanocavity described through its quasinormal modes (QNMs) \cite{lalanne2018light,lalanne2019quasinormal,wu2021nanoscale,wu_modal_2023,dal2025nonlinear}. In a nanocavity, the Green's function is dominated by its resonances. For a mode \(m\) with a complex eigenfrequency \(\tilde{\omega}_m = \omega_m + i\gamma_m\), the retarded dyadic Green’s function near resonance is expanded using the first-order pole form from the Cauchy residue \cite{resStates,resStates2,dal2025nonlinear,kristensen2011effective}:
\begin{equation}
\mathbf{G}(\mathbf{r},\mathbf{r}^{\prime },\omega )=\sum_n\frac{c^{2} \boldsymbol{E}_n(\mathbf{r})\boldsymbol{E}_n(\mathbf{r}^{\prime })}{\tilde{\omega} _n(\omega-\tilde{{\omega }}_n )},
\end{equation}
where $\boldsymbol{E}_n(\vett{r})$ are the vector normalized QNMs of the cavity using \cite{wu_modal_2023, wu_qnmnonreciprocal_resonators_2021,dal2025nonlinear}
\barr\label{eq:QNM_norm}
{\rm QN}_n &=& \int_{\mathcal{D}}\,d^3r\,
\Bigg(
\tilde{\boldsymbol{E}}_n(\mathbf{r}) \cdot \frac{\partial [\omega \varepsilon(\omega)]}{\partial \omega} \tilde{\boldsymbol{E}}_n(\mathbf{r})\nonumber\\
&-&
\tilde{\boldsymbol{H}}_n(\mathbf{r}) \cdot \frac{\partial (\omega \mu/\varepsilon_0)}{\partial \omega} \tilde{\boldsymbol{H}}_n(\mathbf{r})
\Bigg)\nonumber\\
&=& \int_{\mathcal{D}}\,d^3r\,
    \Bigg(
    2\tilde{\boldsymbol{E}}_n(\mathbf{r}) \cdot \frac{\partial [\omega^2 \varepsilon(\omega)]}{\partial (\omega^2)} \tilde{\boldsymbol{E}}_n(\mathbf{r})
    \Bigg),
\earr
where $\tilde{\boldsymbol{E}}_n$ and $\tilde{\boldsymbol{H}}_n$ are the non-normalized complex electric and magnetic QNM fields, $\mathcal{D}=\Omega \cup \Omega_{\text{PML}}$ is the integration domain, with $\Omega$ being the physical space, and $\Omega_{\rm PML}$ the perfectly matched layer (PML). In Eq.~\eqref{eq:QNM_norm}, moreover, we are assuming a non-dispersive magnetic permeability $\mu$.
Consequently, the normailized QNM is given by $\tilde{\boldsymbol{E}}_n / \sqrt{{\rm QN}_n}$.
Is important to notice that the modal volume is proportional to the normalization and in particular we have \cite{wu_modal_2023, wu_qnmnonreciprocal_resonators_2021,dal2025nonlinear}
\begin{equation} \label{eq:mode_volume}
    \tilde{V}_n(\mathbf{r}, \mathbf{p}) =
    \left[
    2 \left( \tilde{\boldsymbol{E}}_n \cdot \mathbf{p} \right)^2 / {\rm QN}_n
    \right]^{-1} = \frac{{\rm QN}_n}{2 \left( \tilde{\boldsymbol{E}}_n \cdot \mathbf{p} \right)^2}\,,
\end{equation}
where $\mathbf{p}$ is the polarization direction.
Using the expansion of the Green function, the trace of the imaginary part of the Green's tensor can be readily calculated as
\begin{equation}
	\operatorname{Tr}\left[\operatorname{Im}[\mathbf{G}(\mathbf{r},\mathbf{r},\omega _p)]\right]
    \approx \frac{2Q_{p}c^{2}}{\omega^{2}_{p}}|\boldsymbol{E}_{p}(\mathbf{r})|^{2},
\end{equation}
where $Q_m=\omega_m/2\gamma_m$ is the cavity quality factor, $\gamma_m\ll\omega_m$, and $\omega_p=\omega_m$ (i.e., the pump field is resonant with the $m$-th QNM of the nanocavity) have been assumed.
In order to proceed in the evaluation of the XPM frequency shift of the cavity, we need to express the probe field in terms of the QNMs of the nanocavity.
Since we have used the normalized QNMs, we write:
\begin{equation}
	|\vett{u}_{s}(\vett{r},\omega_{s})|^{2}=E^{2}_{0}\,\eta^2\,Q_s\,|\boldsymbol{E}_{s}(\mathbf{r})|^{2},
\end{equation}
where $E^{2}_{0}=\hbar\omega/\epsilon_{0}$ is the dimensional constant introduced to recover the correct dimensions of the electric field generally expressed as the product of the single-photon amplitude $\sqrt{{\hbar\omega}/{\epsilon_{0}V_{s}}}$ and the electric field mode function, and the overlap factor between the incident field and the excited QNM is \cite{wu_modal_2023}:
\beq
\eta=\int_V\,d^3r\,\Delta\varepsilon(\vett{r},\omega_s)\boldsymbol{E}_s(\vett{r})\cdot\boldsymbol{E}_{exc}(\vett{r}),
\eeq
where $\Delta\varepsilon(\vett{r},\omega_s) = \varepsilon(\vett{r},\omega_s) - 1$ is the difference between the relative permittivity of the structure and of the surrounding medium (i.e., air in this case) and $\boldsymbol{E}_{exc}(\vett{r})$ is the mode function of the exciting field incident from air.
Therefore, if we insert the above expression in Eq. (\ref{cross}) we obtain:
\begin{equation}\label{Eq.34}
	\Delta\omega_{\text{XPM}}=\Omega_{\text{XPM}}
	\int\,d^3r\,	\chi^{(3)}(\mathbf{r};\omega_{p};\omega_{s})|\boldsymbol{E}_{p}(\mathbf{r})|^{2}|\boldsymbol{E}_{s}(\mathbf{r})|^{2}.
\end{equation}
where $\Omega_{\text{XPM}}=6\Delta\omega_{p}\hbar\omega_{s}\eta^2 c_{\tau} Q_{p}Q_s/2\pi\epsilon_{0}$, where we introduce the attenuation factor that takes into account the loading dynamics of the cavity as well as the temporal overlap of the pump and the signal (see supplementary material):
\begin{equation}
	c_{\tau} = \left(1-e^{-\frac{\Delta t_{p}}{2\tau_p}}\right)^2 \frac{\Delta t_{p}+\tau_p}{\Delta t_{p}+\tau_p +\tau_s}\,.
\end{equation}
It is worth noticing that we work with single-photon pump pulses optimized on the picosecond timescale and a continuous-wave (CW) probe signal in order to optimally balance the contribution of the XPM with respect to the SPM one. This choice also enables maximal interaction with the free-carrier nonlienarities of ENZ media that develop on the picosecond timescale \cite{KinseyReview}.
Equation~\eqref{Eq.34} is the main result of our paper. 
We note that $\Delta\omega_{\text{XPM}}$ is proportional to the multiplication of the two quality factor ($Q_p Q_s$) and in first approximation, due to Eq.~\eqref{eq:mode_volume}, is inversely proportional to the product of the two nanoscale modal volumes [$1/(V_p V_s)$]. Consequently, the ability of designing a nanocavity with high quality factor and small modal volume at both angular frequencies is crucial to obtain a measurable $\Delta\omega_{\text{XPM}}$.

\section{Topology-optimized dual-wavelength silicon-cadmium oxide nanocavity}\label{sec:topopt}

\noindent To achieve an enhanced cross-phase modulation ($\Delta\omega_{\text{XPM}}$) for the detection of a single mid-IR photon ($\lambda_p = 3\,\mu\text{m}$) using a near-IR probe ($\lambda_s = 1.5\,\mu\text{m}$), several demanding optical conditions must be simultaneously met.
The system requires a highly nonlinear material, such as an ENZ medium, embedded within a nanocavity that supports dual resonances at both $\lambda_p$ and $\lambda_s$.
Furthermore, these resonant modes must exhibit high quality factors and deep sub-wavelength modal volumes to maximize light-matter interaction. Unfortunately, conventional design strategies relying on simple canonical geometries are inadequate for this task.
Traditional structures typically support single-wavelength fundamental resonances and lack the degrees of freedom required to achieve sub-wavelength modal volumes at two widely separated wavelengths simultaneously.

To overcome these limitations, we employ topology optimization (TopOpt) \cite{christiansen2021inverse, kang_large-scale_2024, albrechtsen2022nanometer, ma_inverse-designed_2023, albrechtsen2022nanometer,jensen2011topology, xiong2024experimental} to inversely design a dual-wavelength hybrid nanocavity.
As the nonlinear medium, we selected high-mobility cadmium oxide (CdO), which exhibits an ENZ crossing at $\lambda_{\rm ENZ} = 3\,\mu\text{m}$ and a correspondingly high third-order susceptibility.
However, CdO suffers from significant intrinsic optical losses within its ENZ regime.
To mitigate this absorption, we propose a hybrid architecture consisting of a central CdO cylinder with a diameter $D_{\rm ENZ}=20\,\text{nm}$, surrounded by a topology-optimized structure composed of silicon and air. The high refractive index of the surrounding silicon tightly confines the electromagnetic field, minimizing radiation losses and effectively compensating for the dissipation in the ENZ nonlienar region.

Given the immense computational resources required for 3D topology optimization, we approximated the inverse design process using a 2D computational domain.
This choice is often employed in the design of photonic crystal nanocavities via computationally intensive optimization techniques \cite{trojak2020cavity,quan2011deterministic,mignuzzi2019nanoscale,qiu2002effective,painter1999defect}.
To further alleviate the computational burden, we exploited the structural symmetries of the device, simulating only one-quarter of the physical footprint.
To maintain physical accuracy, we utilized the effective refractive index of the TE mode of a standard $220\text{-nm}$-thick silicon slab. It is crucial to note that despite the 2D approximation, the full wavelength-dependent material dispersion for all constituents (Si, CdO) was rigorously incorporated into the simulations.

\begin{figure*}[!t]
    \centering
    \includegraphics[width=\textwidth]{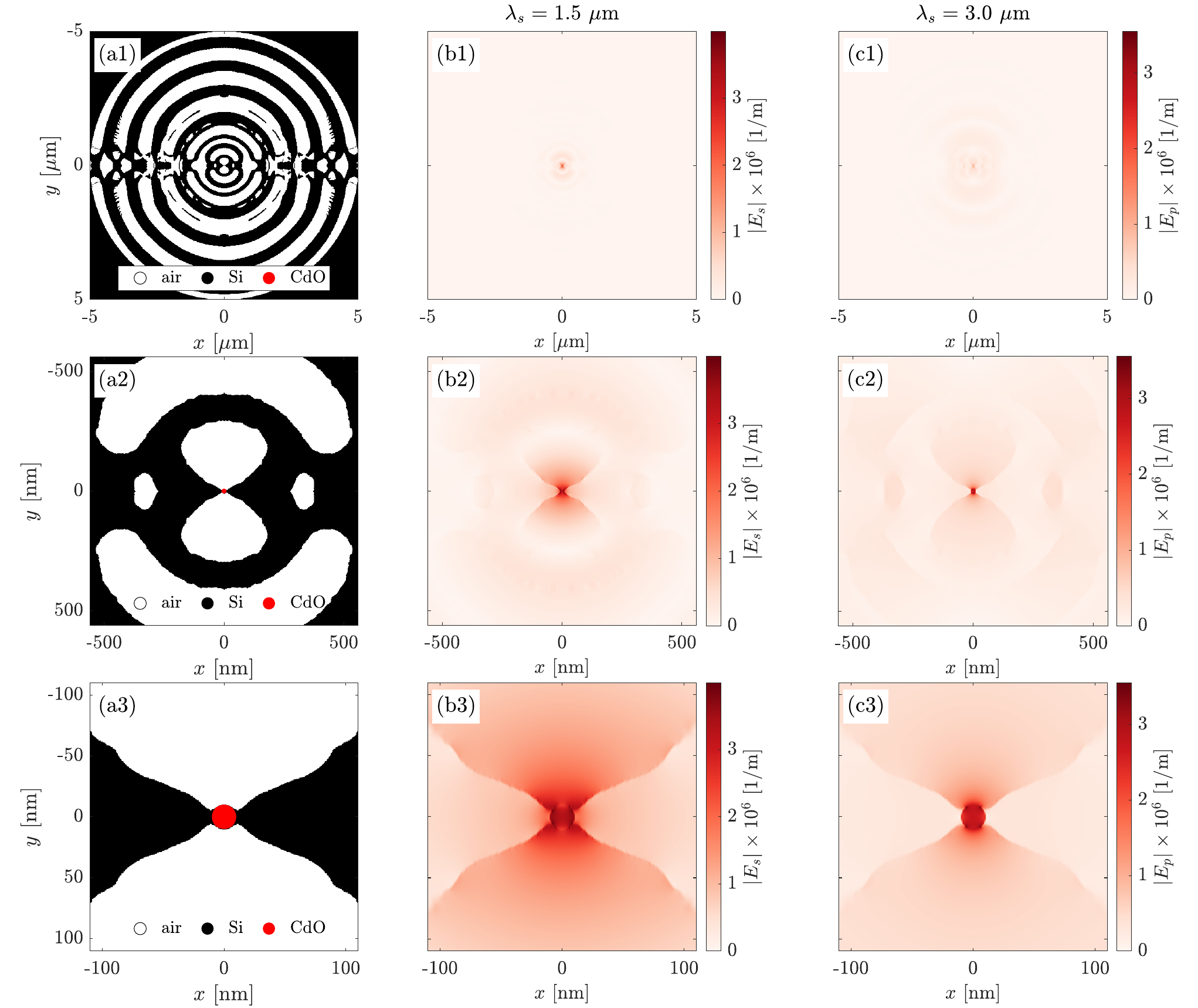}
    \caption{Topology-optimized 2D hybrid nanocavity supporting dual-wavelength resonances. (a1–a3) Material distribution of the inverse-designed cavity structure, composed of air (white), silicon (Si, black), and a central cadmium oxide (CdO, red) cylinder with a diameter of $20\,\rm{nm}$. (b1–b3) Spatial distribution of the electric field magnitude ($|\boldsymbol{E}_s|$) for the QNM at $\lambda_s = 1.5\,\mu\text{m}$. (c1–c3) Electric field magnitude ($|\boldsymbol{E}_p|$) for the QNM at $\lambda_p = 3.0\,\mu\text{m}$. The top row (a1–c1) displays the full optimized footprint, while the middle (a2–c2) and bottom (a3–c3) rows provide progressively magnified views of the central region, highlighting the subwavelength field confinement at the core. The complex angular frequencies of the two modes are $\tilde{\omega}_s \approx 1.2557\,\rm{PHz} + i\,0.13639\,\rm{THz}$ and $\tilde{\omega}_p \approx 0.62768\,\rm{PHz} + i\,0.74837\,\rm{THz}$, yielding quality factors of $Q_s\approx4603$ and $Q_p\approx419$.
    }
    \label{fig:Nanocavity}
\end{figure*}

To the best of our knowledge, this work represents the first demonstration of a topology-optimized nanocavity tailored to operate simultaneously at two widely separated wavelengths. To achieve this, the optimization objective function ($\Phi$) was formulated to maximize the local density of optical states (LDOS).
Specifically, the objective function was defined as the sum of the 2D Purcell factors evaluated at the center of the CdO core for both $\lambda_s$ and $\lambda_p$.
The Purcell factor in a 2D domain is defined as:
\begin{equation}
F_p(\omega)\bigg|_{\rm 2D} = \frac{8}{\omega^2\mu_0}\frac{\Im{\left[E_x(\mathbf{r_0})\right]}}{|p_x|},
\end{equation}
where $E_x$ is the local electric field along $x$, and $p_x$ represents the dipole moment of the emitter. Because the Purcell factor is fundamentally proportional to the ratio $Q_m/V_m$, maximizing this objective function inherently drives the optimizer to discover topologies that simultaneously boost the quality factors and compress the modal volumes to deep sub-wavelength scales.
These terms were appropriately offset and weighted within the objective function to ensure that the optimizer balances the performance enhancement equally across both target wavelengths.
In our TopOpt framework, implemented using the finite element solver COMSOL Multiphysics \cite{comsol_sftw}, the material distribution is governed by a continuous design field $\xi(\mathbf{r}) \in [0,1]$ \cite{christiansen2021inverse}, where $\xi = 0$ corresponds to silicon and $\xi = 1$ to air. The complex refractive index, composed of the real part $n$ and the extinction coefficient $k$, is mapped through a linear interpolation, i.e., 
\begin{equation}
n(\xi(\mathbf{r})) = n_{\text{Si}} + \xi(\mathbf{r}) \left( n_{\text{Air}} - n_{\text{Si}} \right),
\end{equation}
\begin{equation}
k(\xi(\mathbf{r})) = k_{\text{Si}} + \xi(\mathbf{r}) (k_{\text{Air}} - k_{\text{Si}}) + \alpha_a \xi(\mathbf{r})(1 - \xi(\mathbf{r})).
\end{equation}
To prevent the algorithm from converging to non-physical, intermediate refractive indices ($\xi \neq 0,1$), we introduced an artificial attenuation penalty, $\alpha_a \xi(1 - \xi)$, to the imaginary part of the refractive index. This penalization steers the optimizer strictly toward a binary material distribution. Here we use $\alpha_a$ equal to $1$ or $0$.

Finally, to guarantee the practical realizability of the device and comply with state-of-the-art nanofabrication constraints, the raw design field $\xi(\mathbf{r})$ undergoes a two-step processing sequence.
First, a Helmholtz-type differential filter \cite{lazarov_filters_2011} is applied to smooth the field and enforce a minimum feature size, avoiding sharp, unmanufacturable discontinuities. This filtering process is governed by
\beq
- \left( \frac{r_f}{2\sqrt{3}} \right)^2 \nabla^2 \tilde{\xi}(\mathbf{r}) + \tilde{\xi}(\mathbf{r}) = \xi(\mathbf{r}),
\eeq
where $r_f=15\,\rm{nm}$ represents the filter radius and $\tilde{\xi}(\mathbf{r})$ is the resulting smoothed field.
By applying homogeneous Neumann boundary conditions, this step effectively suppresses structural features below the defined length scale.

Subsequently, the filtered field is processed through a smoothed Heaviside projection \cite{christiansen2021inverse} to approximate a discrete binary layout.
The projected physical field $\bar{\xi}(\mathbf{r})$ is then defined as
\beq
\bar{\xi}(\mathbf{r}) = \frac{\tanh(\beta \eta) + \tanh\left( \beta(\tilde{\xi}(\mathbf{r}) - \eta) \right)}{\tanh(\beta \eta) + \tanh\left( \beta(1 - \eta) \right)},
\eeq
where the parameter $\eta \in [0,1]$ defines the transition point (typically set to $0.5$), and $\beta \in [1,\infty)$ controls the steepness of the threshold. To ensure stable convergence, we employed a continuation strategy: the optimization is performed in sequential steps, gradually increasing $\beta \in [2,5,10,50,2000,10^6]$ of the Heaviside projection to drive the structure into a finalized, high-contrast, manufacturable topology where we impose $\alpha_a = 0$.
Once the optimal dual-wavelength hybrid nanocavity was determined, we analyzed its resonant properties using the quasinormal mode (QNM) formalism. To accurately capture the material dispersion and the intrinsic optical losses of the CdO, we incorporated the appropriate weak-form contributions into our computational model to retrieve the correct QNMs, as detailed in  Ref. \cite{wu_modal_2023}.
The analysis successfully yielded two distinct resonant modes at $\lambda_s \approx 1.500\ \mu\text{m}$ and $\lambda_p \approx 3.000\ \mu\text{m}$.
Despite the inherent dissipation characterizing the ENZ regime, the cavity exhibits impressive quality factors of $Q_p \approx 419$ (which is primarily loss-limited by the CdO nonlinear region) for the pump and $Q_s \approx 4603$ for the probe.
Furthermore, via the topology optimization we achieved extreme sub-wavelength light confinement. The integrated modal volumes are $V_p = 1.4 \times 10^{-20}\ \text{m}^{3}$ ($\approx 2 \times 10^{-5} (\lambda_p/n_p)^3$) with $n_p = 0.34$ at the ENZ wavelength, and $V_s = 1.0 \times 10^{-20}\ \text{m}^{3}$ ($\approx 2.3 \times 10^{-2} (\lambda_s/n_s)^3$) with $n_s = 1.99$.
In order to estimate the modal volume, we multiply the modal area evaluated through the 2D simulation by the slab thickness of $220\,\rm{nm}$.
It is worth noticing that the two modal volumes are very similar and that the QNM at $\lambda_p=3\,\mu\text{m}$ is strongly confined inside the CdO nonlinear region despite the fact that the ENZ wavelength is two times larger than the probe one.
This reflects the almost quasi-static character of the field at the ENZ condition, as shown clearly in Fig.~\ref{fig:Nanocavity}.

The resulting optimized geometry and its electromagnetic response are illustrated in Fig.~\ref{fig:Nanocavity}. Panels (a1)–(a3) display the binary material distribution at progressively magnified scales, highlighting the central $20\,\rm{nm}$ CdO nonlinear region surrounded by the Si-air topological layout.
The spatial distributions of the electric field magnitudes, $|\boldsymbol{E}_s|$ (b1–b3) and $|\boldsymbol{E}_p|$ (c1–c3), clearly demonstrate that both QNMs are strongly localized within the central CdO cylinder.
This exceptional spatial overlap at the deep subwavelength scale is a fundamental prerequisite for maximizing the cross-phase modulation efficiency.
The dual-wavelength nanocavity performance is further demonstrated by the Purcell factor spectrum in the supplementary material, which exhibits peak enhancements of $2 \times 10^5$ and $5 \times 10^6$ at $\lambda_s$ and $\lambda_p$, respectively.

\section{Feasibility of quantum nondemolition single-photon detection}\label{feasibility}

\noindent We now apply our formulas to assess the feasibility of mid-IR single-photon detection by the induced XPM phase shift in dual wavelength hybrid nanocavities designed by free-form topology optimization \cite{albrechtsen2022nanometer,jensen2011topology,christiansen2021inverse}.
In order to address this problem, we computed the XPM shift from Eq.~\eqref{Eq.34} using the rigorous formalism of quasi-normal modes for dispersive materials \cite{lalanne2018light,lalanne2019quasinormal,wu2021nanoscale,wu_modal_2023} considering a hybrid silicon nanocavity optimized in the section above.
The CdO materials parameters deduced from the Drude-Sommerfeld dispersion model are consistent with typical experiential data for high-mobility CdO that features a plasma frequency $\omega_{0}=1.45\times 10^{15}$ Hz, a bound permittivity $\epsilon_{b}=5.3$ with an electron collision rate $\gamma=2.8 \times 10^{13}$ Hz. 
Heavily doped transparent conducting oxides like CdO possess a very large electronic third-order susceptibility, particularly when operating near their ENZ frequency, where they are very promising candidates for mid-IR nonlinear device applications \cite{sachet2015dysprosium,schrecengost2024large,reshef2019nonlinear,runnerstrom2017epsilon,vincenti2020enz}.
In what follows we assume a typical XPM third-order susceptibility value of $\chi^{(3)}(\omega_{p};\omega_{s}) \approx 4.4 \times 10^{-17}\,\text{ m}^2/\text{V}^2$ and consider a realistic intensity coupling efficiency $\eta^{2}=0.3$ that is achievable using far-field radiation pattern optimization \cite{albrechtsen2022nanometer,xiong2024experimental,li2026sub}. 
Under these conditions, the designed structure achieves a $\Delta \omega_{\text{XPM}} \approx 3.6\times 10^{10}\ \text{rad/s}$ corresponding to a single-photon cavity pulling \(\Delta\omega_{\text{XPM}} / \omega_s \approx 2.85 \times 10^{-5}\).
The absolute phase shift \(\Delta \theta_s\) accumulated by the classical probe beam is the product of the frequency shift and its average residence (interaction) time \(\tau_{\text{dwell}} = Q_s / \omega_s\approx 3.66\,\text{ps}\) inside the nanocavity. Using our parameters we obtain the phase shift is
\begin{align}
\Delta \theta _{s}&=\Delta \omega_{\text{XPM}}\cdot \tau _{\text{dwell}}=\Delta \omega_{\text{XPM}}\left(\frac{Q_{s}}{\omega _{s}}\right)\\
&=\left(\frac{\Delta \omega_{\text{XPM}}}{\omega _{s}}\right)Q_{s}\approx 0.13\,\,\text{rad}.
\end{align}
A deterministic phase rotation of \(\sim 7.5^{\circ }\) per single mid-IR photon is well above the fundamental resolution bounds of standard homodyne or heterodyne interferometric setups.
Moreover, while the spatial confinement (\(V_{m}\)) drives this giant XPM phase shift, the modest quality factor (\(Q_{p} \sim 419\)) limits the cavity lifetime to \(\tau = Q_p/\omega_p\approx 668\,\text{fs}\).
This indicates that the mid-IR photon wavepacket must be ultrashort to match the high-speed cavity dynamics.

While the obtained single-photon cross-phase modulation (XPM) frequency shift \(\Delta f_s \approx 5.71\text{ GHz}\) and fractional tuning $\Delta\omega_{\text{XPM}}/\delta\omega_{s}=\Delta\omega_{\text{XPM}}Q_{s}/\omega_{s}\approx 0.13$ are extraordinarily large, translating this into a practical, room-temperature quantum nondemolition (QND) mid-IR detector faces major physical limits. The four primary challenges are thermo-refractive noise, intrinsic electronic jitter/thermal carrier extraction, shot noise, and Self-Phase Modulation (SPM), as listed below.
\begin{enumerate}
	\item \textbf{Thermorefractive Noise (TRN) Limit}: thermorefractive noise arises from thermodynamic temperature fluctuations within tiny volumes, which directly modulate the material's refractive index ($\Delta n=(dn/dT)\Delta T$). While compressing the mode volume to \(V_s = 1.0 \times 10^{-20}\text{ m}^3\) boosts the single-photon frequency shift, it penalizes the thermal stability. The variance of local temperature fluctuations, in fact, scales inversely with volume \cite{landau2013statistical,aspelmeyer2014cavity}
	\begin{equation}
		\langle \Delta T^{2}\rangle =\frac{k_{B}T^{2}}{\rho C_{v}V_{m}},
		\end{equation}
		where \(\rho \) is density and \(C_{v}\) is the specific heat. For instance, for a  modal volume of $V_m=10^{-22}\text{ m}^3$ at room temperature, \(\sqrt{\langle \Delta T^{2}\rangle }\) can exceed several degrees Kelvin.
        Similarly to the case of ITO, doped CdO operating near its ENZ is expected to exhibit an extraordinarily large thermo-optic coefficient \cite{wu2024thermo, choi2013temperature} $dn/dT = 2\times10^{-4}{\rm K^{-1}}$ (see supplementary material). However, it is important to realize that in our framework we use a signal frequency far from the ENZ condition and only the nanoscale modal volume contributes to significantly enhance the temperature fluctuations. These massive temperature fluctuations then shift the cavity resonance frequency \cite{gorodetsky2004fundamental,huang2019thermorefractive} by
		\begin{equation}
			\delta \omega _{\text{TRN}}=\frac{\omega _{s}}{n}\frac{dn}{dT}\sqrt{\langle \Delta T^{2}\rangle }.
		\end{equation}
		Therefore, if \(\delta \omega_{\text{TRN}}\) approaches or exceeds the computed single-photon shift ($\Delta \omega_{\text{XPM}} \approx 3.6\times 10^{10}\ \text{rad/s}$), the deterministic shift gets buried in a fluctuating thermal background. To resolve this, the system must either be cryogenically cooled or probed using ultra-short pulses faster than the thermal correlation time (\(\tau_{\text{thermal}} \sim \text{ps}\)).
	\item \textbf{Free-Carrier Absorption Effect}: because CdO is a heavily doped transparent conducting oxide (TCO), its large \(\chi ^{(3)}\) nonlinearity is closely linked to its conduction band plasma dynamics.
		In particular, there are hot electron redistribution effects. This is because when a mid-IR pump photon enters the cavity, it does not just induce an instantaneous bound-electron Kerr shift but it can cause localized absorption that excites hot electrons. These carriers relax via electron-phonon scattering over timescales of \(\sim 100\text{ fs}\) to a few picoseconds.
		In addition, there will also be timing jitter via carrier jittering. The stochastic nature of the arrival time of a photon that enters the cavity versus the times when these hot carriers thermalize or recombine, in fact, create significant timing jitter. If the probe pulse arrival jitters relative to the transient \(\chi ^{(3)}\) window, the experienced phase integration changes from pulse to pulse, adding massive phase noise to the readout circuit.
    \item \textbf{Cavity Linewidth Readout and Shot-Noise}: because the cavity mode volume is so small, its quality factor is limited by the high Ohmic losses inherent to conducting oxides. In particular, a quality factor of $Q_{s}=4603$ at \(\lambda_s = 1.5\,\mu\text{m}\) implies a raw cavity linewidth of:
    \begin{equation}
        \delta\omega_{s}=\frac{\omega _{s}}{Q_{s}}\approx 2.73\times 10^{11}\text{\ rad/s}\,,
    \end{equation}
    which corresponds to $\delta f_{s}\approx 43\text{\ GHz}$.
    Therefore, even though the single-photon frequency shift (\(\Delta f_s \approx 5.71 \text{ GHz}\)) is an appreciable fraction of the linewidth ($\approx 13\%$), the probe spectral profile still broadly overlaps with its shifted state.
    Finally, there will also be shot noise limitations. To distinguish a \(5.71\text{ GHz}\) frequency shift on a broad \(43\text{ GHz}\) line, the classical probe must still contain enough photons (\(N_{s}\)) so that the phase resolution outpaces the quantum projection noise of the probe itself (\(\Delta \theta_{\text{shot}} = 1/(2\sqrt{N_s})\)) \cite{itano1993quantum}.
    However, increasing \(N_{s}\) turns on Self-Phase Modulation (SPM), which feeds back as phase noise and dampens the signal, as we will discuss in Sec.~\ref{SPMeffects}.
    \item \textbf{Self-Phase Modulation}: Self-Phase Modulation (SPM) introduces a critical feedback mechanism that increases phase fluctuations due to the SPM backaction.
    This effects has to be carefully considered in order to establish the feasibility of the approach.
    Because SPM originates from the exact same $\chi^{(3)}$ electronic nonlinearity as XPM, it cannot be turned off.
    The former, in fact, acts as an asymmetric power constraint, and its effect on the single-photon pump is fundamentally different than its effect on the classical probe.
\end{enumerate}
These preliminary considerations point to the necessity of a rigorous feasibility analysis based on experimentally available or theoretically calculated material parameters for the different phase noise contributions.
To quantitatively evaluate the experimental feasibility of single mid-IR photon detection in the proposed CdO hybrid nanocavity, we must optimize the readout state of the classical probe beam.
This requires modeling the total system noise, deriving the combined Signal-to-Noise Ratio (SNR), and calculating the optimal probe photon number (\(N^\ast_{s}\)) that balances fundamental quantum shot noise against probe-induced nonlinear carrier absorption.
Before discussing the modeling of the combined noise and SNR, we introduce the effect of SPM in greater detail below.

\subsection{The Effect of Self-Phase Modulation}\label{SPMeffects}

\noindent In this section, we discuss in detail what is the effect of SPM on the pump and probe beam, and how it interacts with the squeezed probe.

\subsubsection{Effect on the Pump beam}

\noindent For the mid-IR pump beam ($\ket{1}_p$), the expected SPM frequency shift is governed by its own photon number operator $\hat{N}_p$. Since it is strictly a single-photon state, there are no additional photons within that mode to stochastically self-modulate. The expectation value of the pump-pump interaction corresponds only to a static shift of the single-photon vacuum fluctuation baseline, which is fully absorbed into the unperturbed cavity resonance frequency definition. Therefore, the pump wavepacket undergoes zero dynamic SPM distortion. We then conclude that the effect of SPM on the Single-Photon Pump is negligible.

\subsubsection{Effect on the Classical Probe Beam}

\noindent However, because the probe beam is in a classical, coherent state $\ket{\alpha}_s$, this is a much stronger effect, and it is, in fact, the dominant one since in this case SPM acts as a non-linear self-action. Since the probe possesses an typically high average number of photnons, these photons stochastically modulate their own local refractive index, triggering two major limitations:
\begin{enumerate}
    \item[(i)] \textbf{Power-Dependent Cavity Resonance Detuning}: the probe induces a deterministic self-frequency shift ($\Delta\omega_{\text{SPM}}$) on its own cavity mode \cite{drummond1980quantum,walls2008quantum,dal2025nonlinear}
    \begin{equation}
    	\Delta\omega_{\text{SPM}} = \frac{3\hbar\omega_s^2 \chi^{(3)}(\omega_s)}{4\epsilon_0 V_s} N_s.
    \end{equation}
    In the following, we apply Miller's rule [Eq.~\eqref{eq:Miller}] to evaluate $\chi^{(3)}(\omega_s)$ for CdO, obtaining $\chi^{(3)}(\omega_s)\approx\chi^{(3)}(\omega_{\rm ENZ})/253$.
    Comparing $\Delta\omega_{\text{SPM}}$ with the cavity linewidth ($\delta f_{s} \approx 43\text{ GHz}$), the self-shift could physically push the cavity completely off resonance before the single mid-IR pump photon even arrives.
    However, this problem could be mitigated if the cold (un-pumped) cavity is intentionally fabricated with a blue-detuned resonance, such that when the classical probe couples to the cavity, its own SPM dynamically pushes it into the perfect target resonance window.
    \item[(ii)] \textbf{Amplitude-to-Phase Noise Conversion}: because the coherent probe has a Poissonian photon-number distribution, its amplitude fluctuates stochastically by $\Delta N_s = \sqrt{N_s}$. Through SPM, these amplitude fluctuations are instantly converted into a random phase noise 
    \begin{align}
    	\sigma_{\text{SPM}}^2 &= \left( \frac{\partial \Delta\omega_{\text{SPM}}}{\partial N_s} \tau_{\text{dwell}} \right)^2 \Delta N_s^2\nonumber\\
        &= \left( \frac{3\hbar\omega_s^2 \chi^{(3)}(\omega_s)}{4\epsilon_0 V_s}\frac{Q_s}{\omega_s} \right)^2 N_s = b N_s\,,
    \end{align}
    where $b := [ 3\hbar\omega_s \chi^{(3)}(\omega_s) Q_s / (4\epsilon_0 V_s) ]^2$.
    This noise directly competes with the single-photon XPM signal, and need to be considered for the readout fidelity.
\end{enumerate}

\subsection{Modelling the Combined Noise and SNR}

\noindent To read out the cross-phase modulation (XPM) frequency and phase shift (\(\Delta\theta_s \approx 0.13\text{ rad}\)), the cavity is interrogated by a classical probe pulse. The total phase variance \(\sigma_{\rm tot}^{2}\) of the probe beam limiting the detection fidelity is a combination of four independent noise sources
\begin{equation}\label{eq:sigma_tot}
    \sigma_{\rm tot}^{2}=\sigma_{\text{shot}}^{2}+\sigma_{\text{TRN}}^{2}+\sigma_{\text{carrier}}^{2}+\sigma_{\text{SPM}}^2\,,
\end{equation}
where $\sigma_{shot}^2$, $\sigma_{\text{TRN}}^{2}$, $\sigma_{\text{carrier}}^{2}$, and $\sigma_{\text{SPM}}^{2}$ are the phase variance contributions due to shot noise, the thermorefractive noise, the free-carrier absorption, and the SPM, respectively.
$\sigma_{shot}^2$ scales inversely with the number of detected photons \cite{gerry2023introductory,giovannetti2011advances}, i.e.,
\begin{equation}
	\sigma _{\text{shot}}^{2}=\frac{A}{N_{s}},
\end{equation}
According to the Collett-Gardiner input-output formalism \cite{gardiner1985input}, the shot noise of the measured output field depends on the external cavity coupling rate $\kappa_{\text{ext}}$, internal material absorption rate $\kappa_{\text{int}}$, and the total cavity decay rate $\kappa = \kappa_{\text{ext}} + \kappa_{\text{int}}$:
\begin{equation}
		A = \frac{\kappa^2}{16 \kappa_{\text{ext}}^2}
\end{equation}
We focus for simplicity the condition of critical coupling ($\kappa_{\text{ext}} = \kappa_{\text{int}}$), which reduces precisely to the standard quantum boundary value of $A = 0.25$.

The thermorefractive noise contributes with the phase variance $\sigma_{TRN}^2$ imparted during the cavity dwell time \(\tau_{\text{dwell}} = Q_s/\omega_s\) due to thermal volume fluctuations as \cite{panuski2020fundamental,huang2019thermorefractive}
\begin{equation}
	\sigma _{\text{TRN}}^{2}=(\delta \omega _{\text{TRN}}\cdot \tau _{\text{dwell}})^{2}=\left(Q_{s}\frac{1}{n}\frac{dn}{dT}\right)^{2}\frac{k_{B}T^{2}}{\rho C_{v}V_{s}},
\end{equation}
which, using the structural parameters for CdO at room temperature (\(T = 300\text{ K}\)) and \(V_s = 1.0 \times 10^{-20}\text{ m}^3\), gives approximately \(\sigma_{\text{TRN}} \approx 3.1 \times 10^{-3}\text{ rad}\).
Finally, $\sigma^2_{carriers}$ is the probe-induced carrier phase noise. As the classical probe intensity \(N_{s}\) increases, parasitic multi-photon or free-carrier absorption generates hot electrons. This changes the refractive index stochastically by \(\Delta n_{\text{car}} \propto \Delta N_s\). We consider here the resulting phase variance due to free-carrier absorption in the probe beam, which scales with the number of cavity photons \cite{hamerly2015quantum} as
\begin{equation}
	\sigma _{\text{carrier}}^{2}=\beta N_{s},
\end{equation}
where \(\beta \) describes the backaction of the intensity fluctuations of the probe that modulates the material's current density, which feeds back into the laser phase. A closed-form expression for the probe-induced carrier backaction coefficient $\beta$ per probe photon can be  obtained from the solution of the coupled quantum Langevin equations for a Drude dispersive medium (see supplementary material).
Specifically, by coupling the quantum Langevin equations of the cavity field to a free-carrier relaxation equation \cite{aspelmeyer2014cavity,gardiner1985input}, the effective phase noise coupling coefficient $\beta$ (in units of $\text{rad}^2/\text{photon}$) is derived as:
	\begin{equation} \label{beta}
		\beta = \left( \frac{\partial \phi}{\partial N} \right)^2 \frac{R_{\text{gen}}^2}{\gamma_R^2}
	\end{equation}
where $\gamma_R$ represents the hot-carrier thermal relaxation rate back to equilibrium, $R_{\text{gen}} = \xi \kappa_{\text{int}} / V$ is the carrier generation rate per photon, and $\xi$ is the quantum yield of fluctuation, i.e., the number of recombined carriers per photon.
For further details, please refer to the supplementary material.
The phase sensitivity term $\partial \phi / \partial N$ modulates the free carrier density $N$ and alters the real part of the Drude permittivity, thus shifting the cavity frequency $\omega_c$.
This can be evaluated explicitly for a Drude model, yielding the phase susceptibility:
\begin{equation}
    \left|\frac{\partial \phi}{\partial N}\right| = \frac{\omega_c e^2}{\kappa {\varepsilon_{b}} \varepsilon_0 m^* \omega^2}
\end{equation}
where $m^*$ is the carrier effective mass, $\varepsilon_{b}$ is the background permittivity.
Substituting in Eq.~\eqref{beta} we can factor the resulting expression for $\beta$ into intrinsic material properties ($\mathcal{M}_{\text{material}}$) and structural cavity geometry ($\mathcal{G}_{\text{cavity}}$):
\begin{equation}
    \beta = \mathcal{M}_{\text{material}} \cdot \mathcal{G}_{\text{cavity}} = \left( \frac{e^2 \xi}{m^* \gamma_R} \right)^2 \left( \frac{\omega_c}{\omega^2\varepsilon_{b}\,\varepsilon_0\,V_{s}} \frac{\kappa_{\text{int}}}{\kappa}\right)^2
\end{equation}
Considering $\xi=1$ and $\kappa_{\text{int}}/\kappa=0.5$ (i.e., critical coupling) together with the typical materials parameters for CdO \cite{hamerly2015quantum,yang2019high,yang2017femtosecond}, we get $\beta \approx 2.5\times 10^{-8}$ $\text{rad}^2/\text{photon}$ for a nanocavity with a diameter $D_{\rm ENZ}=20\,{\rm nm}$.

Based on this analysis, we then introduce a comprehensive Signal-to-Noise ratio (SNR) for detecting the single-photon induced XPM phase shift as \cite{panuski2020fundamental}
\begin{equation}\label{eq:SNR}
	\text{SNR}=\frac{(\Delta \theta _{s})^{2}}{\sigma _{\rm tot}^{2}}=\frac{(\Delta \theta _{s})^{2}}{\frac{A}{N_{s}}+\sigma _{\text{TRN}}^{2}+\beta N_{s}+b N_{s}}.
\end{equation}
To maximize the SNR, we differentiate the denominator with respect to \(N_{s}\) and set it to zero, which yields the optimum operating point $N^{\ast}_{s} = \sqrt{A/(\beta+b)}$ where quantum shot noise perfectly balance out with the non-linear SPM and carrier noise.
This gives an optimal number $N^{\ast}_{s}\approx 35$ of probe photons.
Operating the readout pulse at this average photon number ensures the maximum possible phase sensitivity without driving the CdO nanocavity into a destructive, carrier-induced absorption regime.
If we substitute the value $N^{\ast}_{s}$ into the expression for $\sigma_{\rm tot}^2$ given by Eq. \eqref{eq:sigma_tot}, we get the visibility threshold for a single mid-IR photon of $\sigma_{\rm tot}\approx 0.12\,{\rm rad}$. Using this value into  Eq.~\eqref{eq:SNR} then gives the peak $\text{SNR}\approx 1.2$.
This shows that, single mid-IR photons are detectable using the proposed approach, if a very small number of probe photons is utilized.
In order to be able to work with larger photon numbers without degrading the SNR, we consider below an amplitude-squeezed probe state into our system noise model.

\begin{figure*}[!t]
    \centering
    \includegraphics[width=\textwidth]{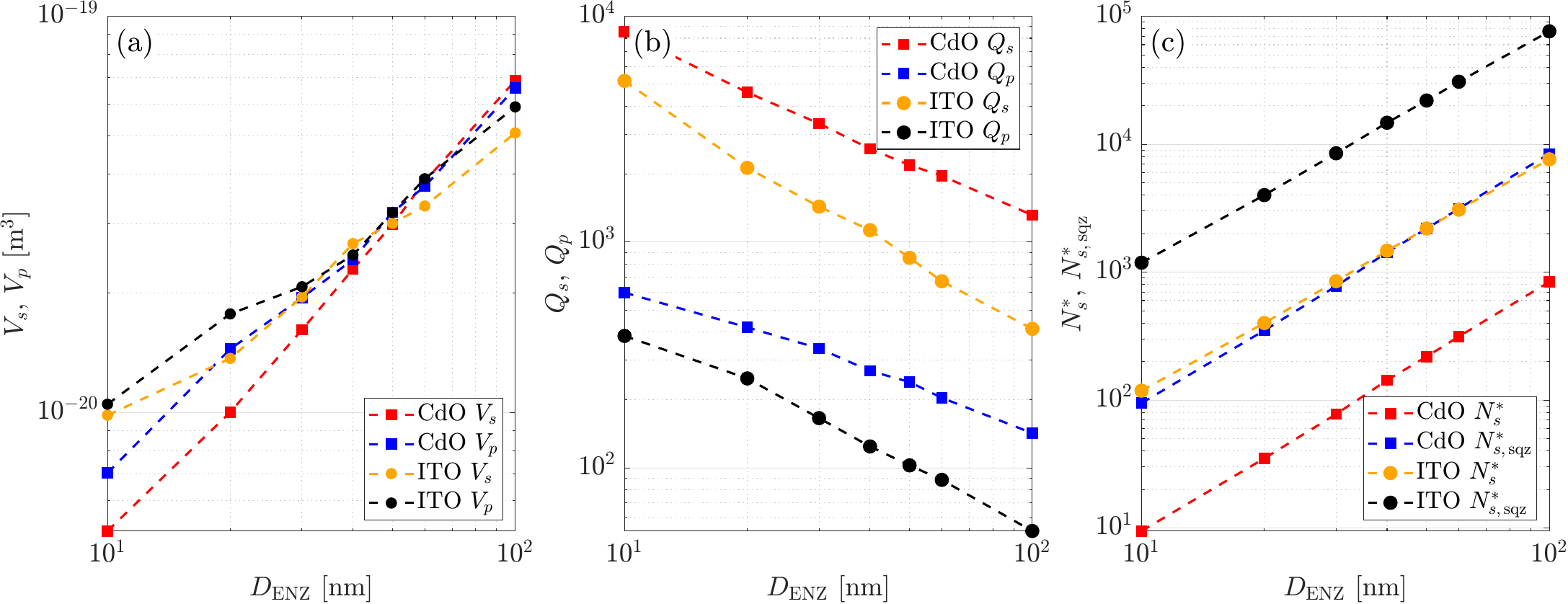}
    \caption{Scaling properties of topology-optimized dual-wavelength nanocavities versus ENZ core diameter $D_{\rm ENZ}$. 
    (a) Modal volumes at the probe ($V_s$) and pump ($V_p$) wavelengths for Si-CdO and SiN-ITO structures. 
    (b) Quality factors of the signal ($Q_s$) and pump ($Q_p$) resonances. 
    (c) Optimal probe photon number required to maximize the signal-to-noise ratio with a classical ($N_s^*$) and a 10 dB amplitude-squeezed ($N_{s,\,{\rm sqz}}^*$) state.}
    \label{fig:V_Q_Ns}
\end{figure*}

\subsection{Amplitude-Squeezed Probe State}

\noindent An amplitude-squeezed coherent state allows us to artificially suppress the quantum noise in the amplitude quadrature below the symmetric shot-noise floor, at the expense of anti-squeezing the phase quadrature \cite{walls2008quantum}.
Let $r$ represent the squeezing parameter, where the squeezing factor in decibels is given by $dB = 10 \log_{10}(e^{2r})$.
The carrier-induced phase noise variance $\sigma_{\text{carrier,\,sqz}}$ decreases proportionally to the variance of the probe photons $\Delta N_s$, modifying the noise term \cite{walls2008quantum,panuski2020fundamental,hamerly2015quantum}, i.e.,
\begin{equation}
	\sigma_{\text{carrier,\,sqz}}^2 \approx \beta N_{s} e^{-2r}.
\end{equation}
In a similar way, due to the fact that $\Delta N_s^2=N_{s} e^{-2r}$, the SPM phase noise also decreases according to:
\begin{equation}
	\sigma_{\text{SPM,\,sqz}}^2 = b N_{s} e^{-2r}.
\end{equation}
However, anti-squeezing increases phase fluctuations, hence the phase variance related to the shot noise becomes \cite{gerry2023introductory}
\begin{equation}
	\sigma_{\text{shot,\,sqz}}^2 = \frac{Ae^{2r}}{N_s}.
\end{equation}
Taking into account this modification, the SNR in the presence of amplitude-squeezing becomes:
\beq\label{eq:SNRsqz}
\text{SNR}_{\text{sqz}} = \frac{(\Delta\theta_s)^2}{\frac{A}{N_s e^{-2r}} + \sigma_{\text{TRN}}^2 + \beta N_s e^{-2r} + b N_s e^{-2r}}.
\eeq
Differentiating again the denominator with respect to $N_s$ considering a fixed squeezing value of $10\text{ dB}$ ($e^{-2r} = 0.1$, $e^{2r} = 10$) leads to the optimal photon number $N^{\ast}_{s,\,\text{sqz}}= e^{2r}\sqrt{A/(\beta+b)}\approx 351\text{\,photons}$.
%Therefore, employing an amplitude-squeezed probe significantly increases the optimal photon number without increasing the total phase noise and altering the SNR.
Therefore, employing an amplitude-squeezed probe significantly increases the optimal photon number without increasing the total phase noise and altering the $\text{SNR}_{\text{sqz}} \approx 1.2$.
This is due to the fact that $N_s$ and $e^{-2r}$ always appear as a product in Eq.~\eqref{eq:SNRsqz}.
However, it is often convenient to work with a probe beam characterized by a larger photon number. For instance, using $N_s = 10\,N^{\ast}_{s,\,\text{sqz}} = 3505$, we achieve $\text{SNR}_{\text{sqz}} \approx 0.24$; however, improving the mode-overlap factor to $\eta^2 = 0.6$ restores $\text{SNR}_{\text{sqz}} \approx 1$.
%will still be able to obtain a satisfactory $\text{SNR}_{\text{sqz}} \approx 8$.

\section{Scaling analysis of XPM-driven single-photon detection in C\lowercase{d}O and ITO nanocavities}\label{sec:scaling}
In order to establish the feasibility of XPM-driven single-photon detection in the proposed hybrid ENZ nanophotonic structures, we extend our study to a number of topology-optimized hybrid dual-wavelength nanocavities with varying diameters of the ENZ cylindrical region $D_{\rm ENZ}=[10,20,30,40,50,60,100]\,{\rm nm}$.

\begin{figure*}[!t]
    \centering
    \includegraphics[width=\textwidth]{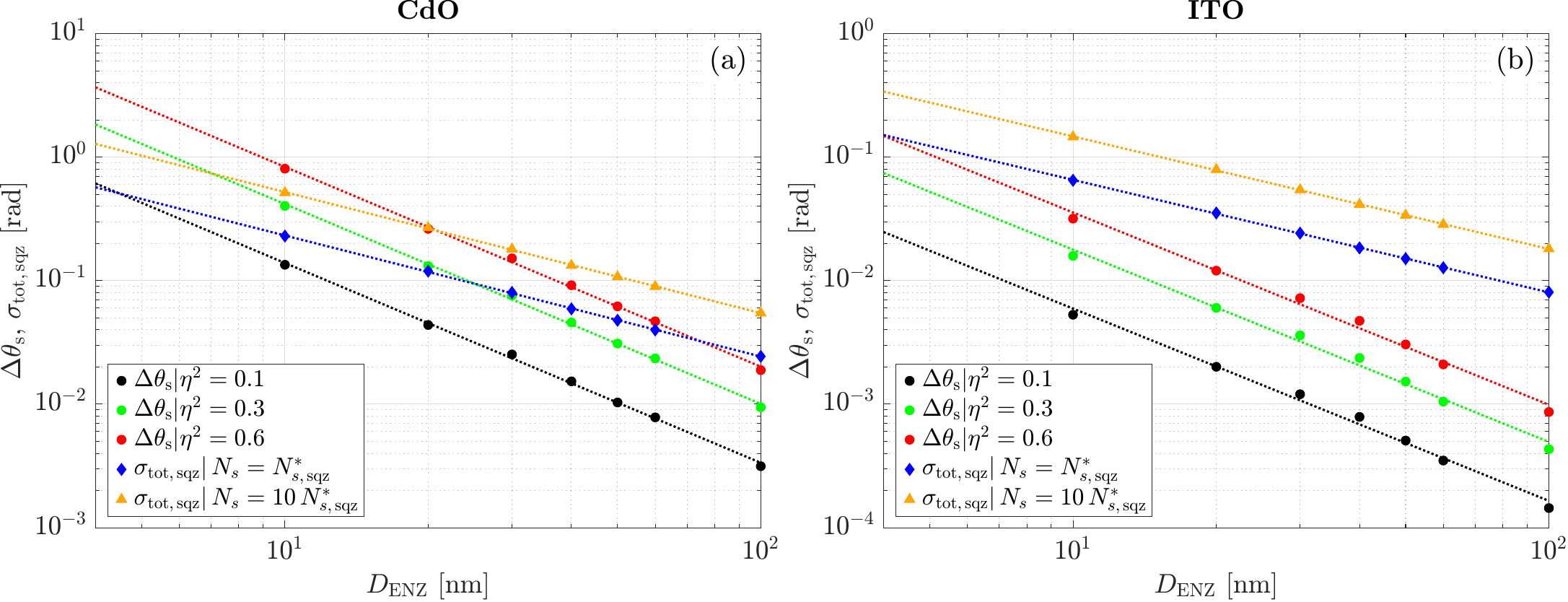}
    \caption{
    Scaling analysis of the single-photon XPM phase shift and competing total phase noise contribution as a function of the ENZ core diameter $D_{\rm ENZ}$.
    Panel (a) reports the results for hybrid Si-CdO topology-optimized dual-wavelength nanocavities, while panel (b) displays the results for hybrid SiN-ITO structures.
    The data compare the deterministic XPM phase shift $\Delta\theta_{s}$ for three different coupling efficiencies ($\eta^2 = 0.1$, $\eta^2 = 0.3$, and $\eta^2 = 0.6$) against $\sigma_{\rm tot,\,sqz}$ for two different average number of probe photons ($N_s = N^{\ast}_{s,\, \rm sqz}$ and $N_s = 10\,N^{\ast}_{s,\,\rm sqz}$).
    The dotted lines indicate power-law fits to the numerical data.
    %The extracted scaling laws for CdO are: $\Delta\theta_s = 9340 \,\eta^2 \, D_{\rm ENZ}^{-2.38}$, $\sigma_{\rm tot,\,sqz}$ equal to $4.27 \, d^{-0.99}$ for $N_s = N^\ast_{s,\, \rm sqz}$ and to $9.58 \, d^{-0.99}$ for $N_s = 10\,N^\ast_{s,\,\rm sqz}$.
    The extracted scaling laws for CdO are: $\Delta\theta_s = 57.7 \,\eta^2 \, D_{\rm ENZ}^{-1.62}$, $\sigma_{\rm tot,\,sqz}$ equal to $2.22 \, D_{\rm ENZ}^{-0.98}$ for $N_s = N^\ast_{s,\, \rm sqz}$ and to $4.98 \, D_{\rm ENZ}^{-0.98}$ for $N_s = 10\,N^\ast_{s,\,\rm sqz}$.
    %For ITO, the fits yield: $\Delta\theta_s = 1050 \,\eta^2 \, D_{\rm ENZ}^{-2.66}$, $\sigma_{\rm tot,\,sqz}$ equal to $2.22 \, d^{-0.92}$ for $N_s = N^\ast_{s,\, \rm sqz}$ and to $4.99 \, d^{-0.92}$ for $N_s = 10\,N^\ast_{s,\,\rm sqz}$.
    For ITO, the fits yield: $\Delta\theta_s = 2.14 \,\eta^2 \, D_{\rm ENZ}^{-1.56}$, $\sigma_{\rm tot,\,sqz}$ equal to $0.534 \, D_{\rm ENZ}^{-0.91}$ for $N_s = N^\ast_{s,\, \rm sqz}$ and to $1.20 \, D_{\rm ENZ}^{-0.91}$ for $N_s = 10\,N^\ast_{s,\,\rm sqz}$.
    In these reported fit parameters, $D_{\rm ENZ}$ is in nanometres and the angles in radians.
    }
    \label{fig:Dtheta}
\end{figure*}

Considering the primary role of the well-characterized ITO material platform for ENZ-enhanced nonlinear optics applications\cite{boyd_nonlinear_2020,vincenti2020enz,capretti2015enhanced,capretti2015comparative,reshef2017beyond, alam2016large, reshef2019nonlinear, dal2025nonlinear}, we also consider here topoly-optimized geometries with $300\text{-nm}$-thick SiN and ITO materials, called SiN-ITO hybrid nanocavities. In this case, since the ENZ wavelength of ITO ($\lambda_{p}=1.234$ $\mu$m) is considerably shorter than the one of CdO, the use of SiN guarantees the complete transmission of the probe beam at $\lambda_{s}=\lambda_p/2=0.617$ $\mu$m.
Detailed spatial profiles and material distributions for the topology-optimized dual-wavelength nanocavities are reported in the supplementary material.
The ITO material platform has been largely investigated in the last few years for the realization of extreme nonlinenar optical responses in the near-infrared spectrum \cite{alam2016large,capretti2015enhanced,capretti2015comparative,shubitidze2024enhanced,shubitidze2025enhancement,reshef2017beyond,reshef2019nonlinear} and refractive index modulations at the single-photon level were recently explored in nanocavity geometries \cite{dal2025nonlinear}.

In what follows, we choose the single-photon pump wavelength to coincide with the ENZ transition wavelength of ITO at $\lambda_{p}=1.234$ $\mu$m and the probe (signal) wavelength to be $\lambda_{s}=\lambda_p/2=0.617$ $\mu$m, while for the Si-CdO hybrid nanocavities we utilize the same pump and signal photon wavelength introduced above. In all our calculations we use the material parameters reported in the supplementary material and we utilize the $\chi^{(3)}(\mathbf{r};\omega_{p};\omega_{s})$ values previously shown in Fig. \ref{fig:chi3_p_s}.

Our results for the scaling of the computed nanocavities'modal volumes, quality factors, and optimal probe photon numbers are reported in Fig. \ref{fig:V_Q_Ns}.
In particular, in Fig. \ref{fig:V_Q_Ns} (a) we show the scaling of the computed modal volumes at the pump ($V_{p}$) and signal ($V_{s}$) wavelengths, which are found to grow linearly by increasing $D_{\rm ENZ}$ with only a small dependence on the type of ENZ material when $D_{\rm ENZ}>40$nm.
The weak dependence of mode volumes from the wavelengths of the pump and signal demonstrates the extreme quasi-static mode localization achieved by topology optimization of dual-wavelength nanocavities.
On the other hand, for the smallest cavity designs ($D_{\rm ENZ}<40$nm), the mode volumes at the pump wavelength appear to be larger than the ones at the signal wavelength due to the rapidly increasing challenges in the optimization of the nanocavities with extremely small ENZ cylindrical regions.
On the other hand, Fig. \ref{fig:V_Q_Ns} (b) features a clear power-law scaling for the quality factors of the nanocavities across the entire range of dimensions $D_{\rm ENZ}$, demonstrating the superior quality of the resonances at both the pump and the probe wavelengths for the CdO-based platform, which is characterized by smaller Ohmic losses compared to its ITO counterpart.
Finally, in Fig. \ref{fig:V_Q_Ns} (c) we show the scaling of the calculated optimal photon numbers considering a classical and an amplitude-squeezed probe beam for both nanocavity materials platforms. Also in this case, our data demonstrate a clear power-law scaling behaviour with varying $D_{\rm ENZ}$ along with significantly reduced cavity photon numbers in the Si-CdO structures compared to the SiN-ITO ones. As already established in the previous section, a small optimal cavity photon number is key to reduce the detrimental effect of SPM phase noise, making Si-CdO structures better candidates for XPM-driven single-photon detection than SiN-ITO ones.

In Fig. \ref{fig:Dtheta} we display our results on the computed XPM single-photon phase shifts for three different values of the coupling efficiency $\eta^{2}$ and the competing total phase noise computed for the optimal average number of probe photons ($N^{\ast}_{s,\,\text{sqz}}$) and also considering $N_s = 10\,N^{\ast}_{s,\,\text{sqz}}$.
It is important to notice that our data demonstrate two very different slopes for the scaling behaviour of the XPM and total phase noise processes versus the ENZ cylinder diameter $D_{\rm ENZ}$.
This provides opportunities to engineer nanocavities with sufficiently small mode volumes so that a crossing point between the $\Delta\theta_{s}$ and the $\sigma_{\rm tot,\, sqz}$ becomes possible for efficiently coupled pump photons.
Figure~\ref{fig:Dtheta} shows in blue colour the $\sigma_{\rm tot,\, sqz}$ for $N_s=N^{\ast}_{s,\,\text{sqz}}$ (also equal to $\sigma_{\rm tot}$ for $N_s=N^{\ast}_{s}$) and in orange colour the $\sigma_{\rm tot,\, sqz}$ for $N_s=10\,N^{\ast}_{s,\,\text{sqz}}$ (also equal to $\sigma_{\rm tot}$ for $N_s=10\,N^{\ast}_{s}$).
Figure~\ref{fig:Dtheta}(a) highlights the importance of a sufficiently high coupling coefficient, $\eta^2 \gtrsim 0.3$, to achieve a SNR greater than $1$ for the hybrid Si-CdO nanocavity with $D_{\rm ENZ} \geq 10\,{\rm nm}$ (whereas for $\eta^2 = 0.1$, the $\text{SNR} \approx 1$ crossing occurs at $D_{\rm ENZ} \approx 4.5\,{\rm nm}$).
The blue curve exhibits a crossing ($\text{SNR} \approx 1$) with the $\eta^2 = 0.3$ case (green curve) around $D_{\rm ENZ} \approx 25\,{\rm nm}$, while is enough $D_{\rm ENZ} \approx 70\,{\rm nm}$ for a stronger coupling of $\eta^2 = 0.6$ (red curve).
Finally, when using a $10\times$ larger probe photon number ($N_s = 10\,N^{\ast}_{s,\,\text{sqz}}$, orange curve) to relax experimental requirements, Fig.~\ref{fig:Dtheta}(a) shows that $\eta^2 = 0.6$ (red curve) is required to obtain the crossing at $D_{\rm ENZ} \approx 20\,{\rm nm}$, whereas it drops to $D_{\rm ENZ} \approx 7\,{\rm nm}$ for $\eta^2 = 0.3$ (green curve).
Overall, Fig.~\ref{fig:Dtheta}(a) demonstrates the feasibility of XPM-driven single-photon detection in hybrid Si–CdO nanocavities, even for $D_{\rm ENZ} \geq 10\,{\rm nm}$.
In particular, using a hybrid Si-CdO nanocavity with $D_{\rm ENZ} = 10\,{\rm nm}$, which could potentially be realized using state-of-the-art nanofabrication techniques\cite{xiong2024experimental, albrechtsen2022nanometer, ouyang2024singular}, we obtain SNR values of $3.1$, $12$, and $2.4$ for the green--blue, red--blue, and red--orange curve combinations, respectively.
In contrast, the results for the hybrid SiN-ITO nanocavities shown in Fig. \ref{fig:Dtheta} (b) do not feature any crossing points in the investigated range of parameters, preventing single-photon detection due to the dominant effect of the SPM noise contribution.
In summary, our comprehensive analysis has established that the total phase noise is an unsurmountable limit for ITO-based nanocavities but can be overcome using hybrid Si-CdO dual-wavelength nanocavities with sufficiently small modal volumes that can potentially be demonstrated using state-of-the-art nanofabrication techniques \cite{xiong2024experimental, albrechtsen2022nanometer, ouyang2024singular} and by integrating high-mobility CdO nanomaterials \cite{sachet2015dysprosium,gulino2003large} in hybrid nanocavity structures.

\section{Conclusions}\label{conclusions}

\noindent In this work, we have used the Green's tensor quantization theory to investigate single-photon detection via enhanced XPM in ENZ nanomaterials confined in the proposed dual-wavelength optical nanocavities designed by free-form topology optimization for both the case of hybrid Si-CdO and SiN-ITO nanocavities.
We derived generally valid closed-form analytical expressions for the achievable XPM phase shift considering in both cases a pulsed single-photon at the ENZ transition wavelength $\lambda_p=\lambda_{\rm ENZ}$ of the material and a CW classical or a squeezed probe at $\lambda_s=\lambda_{\rm ENZ}/2$, both resonantly confined within the same nanoscale volume.
Numerical calculations based on the framework of quasi-normal modes demonstrate the superior quality of the resonances of the Si-CdO platform compared to the SiN-ITO one, enabling a single-photon XPM frequency shift $\Delta f_s \approx 5.71 \text{ GHz}$ with cavity pulling \(\Delta f_s / f_s \approx 2.85 \times 10^{-5}\) for a nanocavity with $D_{\rm ENZ} = 20\,{\rm nm}$.
Finally, we systematically addressed the feasibility of the proposed detection scheme by investigating the practical limitations arising from self-phase modulation noise, thermorefractive noise, shot noise, and free-carrier absorption.
Crucially, we numerically demonstrated the feasibility of XPM-driven single-photon detection in a dual-wavelength hybrid Si-CdO nanocavity, even for $D_{\rm ENZ} \geq 10\,{\rm nm}$. Notably, for $D_{\rm ENZ} = 10\,{\rm nm}$, we achieved an ${\rm SNR} \approx 12$ assuming $\eta^2 = 0.6$, which would required far-field radiation pattern optimization\cite{albrechtsen2022nanometer,xiong2024experimental,li2026sub} in order to be experimentally demonstrated.
This work establishes a robust benchmark for the engineering of mid-infrared single-photon nonlinear devices such as nondemolition quantum detectors, sensors, and all-optical gates on a solid state photonic platform.

\section*{Supplementary Material}
See the supplementary material for additional details regarding the derivation of the carrier backaction coefficient based on the coupled quantum Langevin equations. Additionally, we display all the material's parameters used in the calculations and provide detailed spatial profiles and materials distributions for all the topology-optimized dual-wavelength nanocavities analyzed in this work.

\begin{acknowledgments}
L.D.N. acknowledges the support from the U.S. Army Research Office under Award No. W911NF2510284. M. O. acknowledges the financial support from the Research Council of Finland's Flagship Program (PREIN - decision Grant No. 320165).
L.D.N. and R.F.  are pleased to acknowledge that the computational work reported in this paper was performed on the Shared Computing Cluster, which is administered by Boston University’s Research Computing Services.
L.D.N. acknowledges Professors Iacopo Carusotto and Lorenzo Pavesi for insightful discussions.
\end{acknowledgments}

\section*{Data Availability Statement}
The data that support the findings of
this study are available from the
corresponding author upon reasonable
request.

\bibliography{references}% Produces the bibliography via BibTeX.

% Stampa la bibliografia di questo documento usando il file "references.bib"
% Nota: non serve scrivere l'estensione .bib
%\putbib[references] 

%\end{bibunit} % <-- Fine della prima unità

% ==========================================
%               SEPARATORE
% ==========================================
\clearpage
\setcounter{section}{0}  
\setcounter{figure}{0}   
\setcounter{table}{0}    
\setcounter{equation}{0} 

% --- AGGIUNTA DEL PREFISSO "S" ---
% Ridefinisce il formato dei contatori per il secondo documento
\renewcommand{\thefigure}{S\arabic{figure}}     % Figure diventeranno S1, S2...
\renewcommand{\thetable}{S\arabic{table}}       % Tabelle diventeranno S1, S2...

% Opzionale: puoi farlo anche per le equazioni e per le sezioni
\renewcommand{\theequation}{S\arabic{equation}} % Equazioni diventeranno (S1), (S2)...
\renewcommand{\thesection}{S\arabic{section}}   % Sezioni diventeranno S1, S2...

% ==========================================
%            SECONDO DOCUMENTO
% ==========================================

%\preprint{AIP/123-QED}

%\title{Supplementary Material: Single-Photon Detection via Enhanced Cross-Phase Modulation in Topology-Optimized Epsilon-Near-Zero Dual-Wavelength Nanocavities}
%
%\author{Luca Dal Negro}
% \email{dalnegro@bu.edu.}
%\affiliation{Department of Electrical \& Computer Engineering, Boston University, 8 Saint Mary's Street, Boston, 02215, MA, USA}
%\affiliation{Department of Physics, Boston University, 590 Commonwealth Avenue, Boston,02215, MA, USA}
%\affiliation{Division of Materials Science \&  Engineering, Boston University, 15 St. Mary’s street, Brookline, 02446, MA, USA}
%\author{Riccardo Franchi}%
%\affiliation{Department of Electrical \& Computer Engineering, Boston University, 8 Saint Mary's Street, Boston, 02215, MA, USA}
%\author{Marco Ornigotti}
% \affiliation{Faculty of Engineering and Natural Sciences, Tampere University, Tampere, Finland} 
%
%\date{\today}
%
%\maketitle

% --- TITOLO SUPPLEMENTARY SIMULATO STILE REVTeX ---
\onecolumngrid % Centra su tutta la pagina
\vspace*{1cm}
\begin{center}
    % Titolo
    \textbf{\Large Supplementary Material: Single-Photon Detection via Enhanced Cross-Phase Modulation in Topology-Optimized Epsilon-Near-Zero Dual-Wavelength Nanocavities}\\[0.6cm]
    
    % Autori con richiami ad apice
    Luca Dal Negro$^{1, 2, 3, *}$, Riccardo Franchi$^{1}$, and Marco Ornigotti$^{4}$\\[0.4cm]
    
    % Affiliazioni in corsivo come da standard AIP
    {\small \itshape
    $^1$Department of Electrical \& Computer Engineering, Boston University, 8 Saint Mary's Street, Boston, 02215, MA, USA\\
    $^2$Department of Physics, Boston University, 590 Commonwealth Avenue, Boston, 02215, MA, USA\\
    $^3$Division of Materials Science \& Engineering, Boston University, 15 St. Mary's street, Brookline, 02446, MA, USA\\
    $^4$Faculty of Engineering and Natural Sciences, Tampere University, Tampere, Finland
    }\\[0.4cm]
    
    % Email corrispondente
    {\small $^*$ Email: dalnegro@bu.edu}
\end{center}
\vspace*{1cm}
\twocolumngrid % Torna al formato a due colonne (se usi l'opzione 'reprint')
% --------------------------------------------------

%\begin{bibunit} % <-- Inizia la seconda unità bibliografica

\section{Derivation of the Carrier Backaction Coefficient $\beta$}

In the main text, the probe-induced carrier phase noise variance $\sigma_{\text{carrier}}^2$ is quantified by the backaction coefficient $\beta$, such that $\sigma_{\text{carrier}}^2 = \beta N_s$. This term captures the stochastic refractive index modulation driven by free-carrier absorption. In this section, we provide the rigorous quantum mechanical derivation of $\beta$ starting from the coupled quantum Langevin equations. This first-principles approach reveals the exact physical origins of the phase susceptibility $\partial\phi/\partial N$, the carrier generation rate $R_{\text{gen}}$, and the carrier relaxation rate $\gamma_R$.

\subsection{Coupled Quantum Langevin Equations}

We model the interaction between the fluctuating cavity field, the material's dynamic carrier density, and the phase quadrature of the output laser beam. Consider a nanocavity driven by an incident classical probe laser of frequency $\omega_0$. Let $\hat{a}$ be the annihilation operator of the cavity mode, governed by a total loss rate $\kappa = \kappa_{\text{ext}} + \kappa_{\text{int}}$. In a frame rotating at the laser frequency $\omega_0$, the quantum Langevin equation for the cavity field mode is given by \cite{aspelmeyer2014cavity}:
\begin{equation}
    \frac{d}{dt}\hat{a}(t) = -\left[i\Delta_0 + \frac{\kappa}{2}\right]\hat{a}(t) - i \hat{\Delta}_{\text{carrier}}(t)\hat{a}(t) + \sqrt{\kappa_{\text{ext}}}\hat{a}_{\text{in}}(t)
\end{equation}
where $\Delta_0 = \omega_c^{(0)} - \omega_0$ is the cold cavity detuning, and $\hat{a}_{\text{in}}(t)$ is the incident probe field operator containing the vacuum fluctuations $\hat{\xi}_{\text{in}}(t)$. 

The dynamic frequency shift operator $\hat{\Delta}_{\text{carrier}}(t)$ describes how changes in the free-carrier population $\hat{N}(t)$ alter the cavity resonance via the Drude dispersion. We linearize this shift around the steady-state carrier density $\langle N \rangle$ via a dispersive frequency-pull parameter:
\begin{equation}
    \hat{\Delta}_{\text{carrier}}(t) = \frac{\partial \omega_c}{\partial N} \delta\hat{N}(t)
\end{equation}
where $\delta\hat{N}(t) = \hat{N}(t) - \langle N \rangle$. 

The macroscopic phase noise originates from the stochastic dynamics of this carrier population. The population fluctuations are driven by the absorption of cavity photons and relax back to equilibrium via hot-carrier and phonon scattering pathways. This is described by the rate equation:
\begin{equation}
    \frac{d}{dt}\delta\hat{N}(t) = -\gamma_R \delta\hat{N}(t) + R_{\text{gen}}\hat{a}^\dagger(t)\hat{a}(t) + \hat{F}_N(t)
\end{equation}
Here, we clearly identify two critical physical parameters. The term $\gamma_R$ is the phenomenological hot-carrier thermal relaxation rate. The term $R_{\text{gen}} = \xi \kappa_{\text{int}} / V_m$ is the carrier generation rate per photon, which depends directly on the internal material absorption rate ($\kappa_{\text{int}}$), the modal volume ($V_m$), and the quantum yield of fluctuation ($\xi$). Lastly, $\hat{F}_N(t)$ represents the intrinsic material Langevin noise source.

\subsection{Linearization and Fourier-Domain Analysis}

To isolate the noise components, we decompose the cavity field into its classical steady-state amplitude and a small quantum fluctuation operator: $\hat{a}(t) = \alpha + \delta\hat{a}(t)$, where $\langle n \rangle = |\alpha|^2 = N_s$ is the mean cavity photon number. Assuming the laser operates on-resonance with the steady-state, pulled cavity ($\Delta_0 + \langle \hat{\Delta}_{\text{carrier}} \rangle = 0$), the linearized, coupled fluctuations simplify to:
\begin{equation}
    \frac{d}{dt}\delta\hat{a}(t) = -\frac{\kappa}{2}\delta\hat{a}(t) - i \alpha \left(\frac{\partial \omega_c}{\partial N}\right) \delta\hat{N}(t) + \sqrt{\kappa_{\text{ext}}}\hat{\xi}_{\text{in}}(t)
\end{equation}
\begin{equation}
    \frac{d}{dt}\delta\hat{N}(t) = -\gamma_R \delta\hat{N}(t) + R_{\text{gen}}\alpha\left(\delta\hat{a}(t) + \delta\hat{a}^\dagger(t)\right) + \hat{F}_N(t)
\end{equation}

To find the steady-state phase imprint, we transform these equations into the Fourier domain ($d/dt \to -i\Omega$). Solving the linearized carrier equation yields:
\begin{equation}
    \delta\hat{N}(\Omega) = \frac{R_{\text{gen}}\alpha \left[\delta\hat{a}(\Omega) + \delta\hat{a}^\dagger(\Omega)\right] + \hat{F}_N(\Omega)}{\gamma_R - i\Omega}
\end{equation}
In the low-frequency limit ($\Omega \to 0$), which governs long-lived phase noise and baseline sensor drift, the dynamic response simplifies to:
\begin{equation}
    \delta\hat{N}(0) \approx \frac{R_{\text{gen}}\alpha}{\gamma_R}\left[\delta\hat{a}(0) + \delta\hat{a}^\dagger(0)\right] + \frac{\hat{F}_N(0)}{\gamma_R}
\end{equation}
This highlights the fundamental backaction mechanism: fluctuations in the cavity field amplitude ($\delta\hat{a} + \delta\hat{a}^\dagger$) directly drive the steady-state carrier density fluctuations.

To connect the internal cavity fluctuations to an external observable, we use the Collett-Gardiner input-output relation \cite{gardiner1985input}. The field exiting the cavity through the collection port is given by $\hat{a}_{\text{out}}(t) = \hat{a}_{\text{in}}(t) - \sqrt{\kappa_{\text{ext}}}\hat{a}(t)$. For a bright probe field ($\alpha \gg 1$), the phase fluctuation operator $\delta\hat{\phi}_{\text{out}}$ corresponds directly to the phase quadrature of the output field:
\begin{equation}
    \delta\hat{\phi}_{\text{out}}(\Omega) = \frac{\delta\hat{a}_{\text{out}}(\Omega) - \delta\hat{a}_{\text{out}}^\dagger(\Omega)}{2i \alpha_{\text{out}}}
\end{equation}
Evaluating the field fluctuation transformations relates the internal cavity frequency changes directly to the output phase shift:
\begin{equation}
    \delta\hat{\phi}_{\text{out}}(0) = \delta\hat{\phi}_{\text{shot}}(0) + \left(\frac{\partial \phi}{\partial N}\right) \delta\hat{N}(0)
\end{equation}
where the term $\partial\phi/\partial N$ naturally emerges as the total phase susceptibility. It represents how the dispersive internal frequency pull ($\partial \omega_c/\partial N$) translates into a measurable phase shift at the output port, filtered by the cavity lifetime:
\begin{equation}
    \frac{\partial \phi}{\partial N} = \frac{\partial \phi}{\partial \omega_c} \frac{\partial \omega_c}{\partial N} = \left(\frac{2}{\kappa}\right)\frac{\partial \omega_c}{\partial n}\frac{\partial n}{\partial N}
\end{equation}

Using the binomial expansion we can approximate the Drude refractive index at shorter wavelength with respect to for $\lambda_{\rm ENZ}$ and for $\omega\gg \gamma$ as:
\begin{equation}
    n \approx \sqrt{\varepsilon_{b}} - \frac{N e^2}{2\sqrt{\varepsilon_{b}}\varepsilon_0 m^* \omega^2}
\end{equation}
Using these approximation we obtain:
\begin{align}
    \frac{\partial n}{\partial N} &\approx - \frac{e^2}{2\sqrt{\varepsilon_{b}}\varepsilon_0 m^* \omega^2}\\
    \frac{\partial \omega_c}{\partial n} &\approx -\frac{\omega_c}{\sqrt{\varepsilon_{b}}}\\
    \left|\frac{\partial \phi}{\partial N}\right| &= \frac{\omega_c e^2}{\kappa \varepsilon_b \varepsilon_0 m^* \omega^2}
\end{align}

Taking the statistical variance ($\langle \delta\hat{\phi}_{\text{out}}^2 \rangle$) of our output phase relation reveals two uncorrelated components (the quantum shot noise bath and the intensity-driven carrier loop):
\begin{equation}
    \Delta \phi^2_{\text{total}} = \Delta \phi^2_{\text{shot}} + \left(\frac{\partial \phi}{\partial N}\right)^2 \langle \delta\hat{N}(0)^2 \rangle
\end{equation}
By computing the auto-correlation spectral density of the probe-driven carrier terms $S_{NN}(0) = \langle \delta\hat{N}(0)^2 \rangle$ from our resolved Fourier relation, we obtain:
\begin{equation}
    \langle \delta\hat{N}(0)^2 \rangle = \frac{R_{\text{gen}}^2 \alpha^2}{\gamma_R^2} \langle \left(\delta\hat{a}(0) + \delta\hat{a}^\dagger(0)\right)^2 \rangle = \frac{R_{\text{gen}}^2 \langle n \rangle}{\gamma_R^2}
\end{equation}
Substituting this carrier spectral density back into the phase variance formula yields the total phase noise in terms of the mean photon number $N_s = \langle n \rangle$:
\begin{equation}
    \Delta \phi^2_{\text{total}} = \frac{A}{\langle n \rangle} + \left[\left(\frac{\partial \phi}{\partial N}\right)^2 \frac{R_{\text{gen}}^2}{\gamma_R^2}\right] \cdot \langle n \rangle
\end{equation}
Extracting the bracketed multiplier isolates the exact expression for the backaction coefficient utilized in the main text:
\begin{equation}
    \beta = \left(\frac{\partial \phi}{\partial N}\right)^2 \frac{R_{\text{gen}}^2}{\gamma_R^2} \quad \left[\text{rad}^2/\text{photon}\right]
\end{equation}
This completes the mathematical proof that $\beta$ stems directly from the internal quantum coupling of the optical field amplitude to the free-carrier Langevin population bath.

\section{Temporal Overlap Dynamics and Cavity Loading Equations}
\subsection{Derivation of Internal Cavity Time Dynamics via TCMT}
To rigorously characterize the time evolution of the internal field, we model the system using Temporal Coupled-Mode Theory (TCMT). Let $\alpha_m(t)$ denote the internal cavity mode amplitude of mode $m \in \{p, s\}$ (normalized such that $|\alpha_m(t)|^2$ is proportional to the circulating intensity $I_m(t)$), and let $E_{\text{in}, m}(t)$ be the driving field incident at the cavity.

The time-evolution of the internal mode amplitude is readely described by the following TCMT equation for the classical mode amplitude:
\begin{equation}
    i\frac{d\alpha_m(t)}{dt} = (\omega_m - i\gamma_{\text{tot}, m})\alpha_m(t) - \sqrt{2\Gamma_m}\, E_{\text{in}, m}(t)\,,
\end{equation}
where $\gamma_{\text{tot}, m} = \gamma_m + \Gamma_m = \omega_m/(2 Q_m)$ is the total field amplitude decay rate, $\gamma_m$ is the intrinsic amplitude loss rate (material absorption, scattering, and radiation leakage) and $\Gamma_m$ is the extrinsic port coupling amplitude rate into the injection port.
% and the output field is related to the input and internal fields via:
% \begin{equation}
%     E_{\text{out}, m}(t) = E_{\text{in}, m}(t) + i\sqrt{2\Gamma_m}\, \alpha_m(t)
% \end{equation}
Introducing the slowly-varying envelope decomposition $E_{\text{in}, m}(t) = \varepsilon_{\text{in}, m}(t)e^{-i\omega t}$ and $\alpha_m(t) = a_m(t)e^{-i\omega t}$, the envelope evolution equation becomes:
\begin{equation}
    \frac{da_m(t)}{dt} = \left[i\Delta\omega_m - \gamma_{\text{tot}, m}\right] a_m(t) + i\sqrt{2\Gamma_m}\, \varepsilon_{\text{in}, m}(t)
\end{equation}
where $\Delta\omega_m = \omega - \omega_m$ is the angular frequency detuning. On exact resonance ($\Delta\omega_m = 0$):
\begin{equation}
    \frac{da_m(t)}{dt} = -\gamma_{\text{tot}, m} a_m(t) + i\sqrt{2\Gamma_m}\, \varepsilon_{\text{in}, m}(t)
\end{equation}

To capture the loading phase and subsequent cavity ring-down cleanly, we consider a synchronized rectangular flat-top input pulse for both modes that turns on at $t = 0$ and terminates at $t = \Delta t_m$:
\begin{equation*}
    \varepsilon_{\text{in}, m}(t) = \varepsilon_{0, m}\left[\Theta(t) - \Theta(t - \Delta t_m)\right] = \begin{cases} \varepsilon_{0, m}, & 0 \le t \le \Delta t_m \\ 0, & \text{otherwise} \end{cases}
\end{equation*}

\subsubsection{Phase I: The Loading Phase ($0 \le t \le \Delta t_m$)}
Integrating the differential equation with the initial condition $a_m(0) = 0$:
\begin{equation}
    \int_0^{a_m(t)} \frac{da_m}{i\sqrt{2\Gamma_m}\,\varepsilon_{0, m} - \gamma_{\text{tot}, m} a_m} = \int_0^t dt'
\end{equation}
Evaluating this via separation of variables yields the internal cavity mode amplitude buildup:
\begin{equation}
    a_m(t) = \frac{i\sqrt{2\Gamma_m}\,\varepsilon_{0, m}}{\gamma_{\text{tot}, m}} \left(1 - e^{-t / (2\tau_m)}\right)\,,
\end{equation}
where $\tau_m = Q_m/\omega_m=1/(2\gamma_{\text{tot}, m})$ is the cavity photon lifetime.
Since optical intensity $I_m(t)$ scales as the squared modulus of the mode amplitude ($I_m \propto |a_m|^2$), the internal intensity loading profile is given by:
\begin{equation}
    I_m(t) = I_{\text{ss}, m} \left(1 - e^{-t / (2\tau_m)}\right)^2 \quad \text{for } 0 \le t \le \Delta t_m
\end{equation}
where $I_{\text{ss}, m} = \frac{2\Gamma_m}{\gamma_{\text{tot}, m}^2} I_{\text{in}, m} = \frac{2\Gamma_m}{(\gamma_m + \Gamma_m)^2} I_{\text{in}, m}$ is the steady-state intracavity intensity enhancement.

\subsubsection{Phase II: The Decay / Ring-Down Phase ($t > \Delta t_m$)}
When the incident driving pulse terminates at $t = \Delta t_m$, the internal field has reached $a_m(\Delta t_m)$. For all subsequent times $t > \Delta t_m$, the external driving term drops to zero, leaving pure exponential cavity ring-down:
\begin{equation}
    \frac{da_m(t)}{dt} = -\gamma_{\text{tot}, m} a_m(t)
\end{equation}
\begin{equation}
    \implies a_m(t) = a_m(\Delta t_m)e^{-\gamma_{\text{tot}, m}(t - \Delta t_m)}
\end{equation}
Squaring the modulus yields the internal intensity decay profile (for $t > \Delta t_m$):
\begin{equation}
    I_m(t) = I_m(\Delta t_m)e^{-2\gamma_{\text{tot}, m}(t - \Delta t_m)} = I_m(\Delta t_m)e^{-(t - \Delta t_m) / \tau_m}
\end{equation}
where the peak intensity achieved at the end of the loading window is:
\begin{equation}
    I_m(\Delta t_m) = I_{\text{ss}, m} \left(1 - e^{-\Delta t_m / (2\tau_m)}\right)^2
\end{equation}

\subsubsection{Complete Internal Piecewise Intensity Equations}
The complete time-domain intensity profile inside the cavity at the pump wavelength is:
\begin{equation}
    I_p(t) = \begin{cases} I_{\text{ss}, p} \left(1 - e^{-t / (2\tau_p)}\right)^2, & 0 \le t \le \Delta t_p \\ I_p(\Delta t_p) \, e^{-(t - \Delta t_p) / \tau_p}, & t > \Delta t_p \end{cases}
\end{equation}

Therefore, when we consider the internal intensity at the pump wavelength we need to take into account also the charging dynamics and in particular $a_\tau :=\left(1 - e^{-\Delta t_p / (2\tau_p)}\right)^2$ that depends on $\Delta t_p$ and $\tau_p$.
Please note that here we are not considering the field enhancement factor since it is already taken into account in Eqs. (28) and (29) of the main text.
Additionally, the effective duration spent by the pump pulse in the nanocavity can be approximated as the sum of its raw incident duration and its cavity photon lifetime:
\begin{equation}
    T_{\text{eff}, p} \approx \Delta t_p + \tau_p \,.
\end{equation}

\subsection{Effective Pump–Probe Interaction Time}
The effective interaction time between the single-photon pump pulse and the continuous-wave (CW) probe signal within the nanocavity is governed by the effective residence time of the pump pulse in the cavity and the cavity photon lifetime at the signal wavelength, according to:
\begin{equation}
	T_{\rm interaction} = \frac{\Delta t_{p}+\tau_p}{\Delta t_{p}+\tau_p +\tau_s} \tau_s = b_\tau \tau_s\,,
\end{equation}
where $b_\tau := (\Delta t_{p}+\tau_p)/(\Delta t_{p}+\tau_p +\tau_s)$ is the temporal overlap factor ($b_\tau \approx 0$ when $T_{\text{eff}, p}\ll \tau_s$ and $b_\tau \approx 1$ when $T_{\text{eff}, p}\gg \tau_s$).

Therefore, considering the total effect of the cavity loading dynamics together with the pump-probe interaction time, we obtain the attenuation factor:
\begin{equation}
	c_{\tau} = a_{\tau} \, b_{\tau} = \left(1-e^{-\frac{\Delta t_{p}}{2\tau_p}}\right)^2 \frac{\Delta t_{p}+\tau_p}{\Delta t_{p}+\tau_p +\tau_s}\,.
\end{equation}

\begin{figure}[!t]
    \centering
    \includegraphics[width=0.95\linewidth]{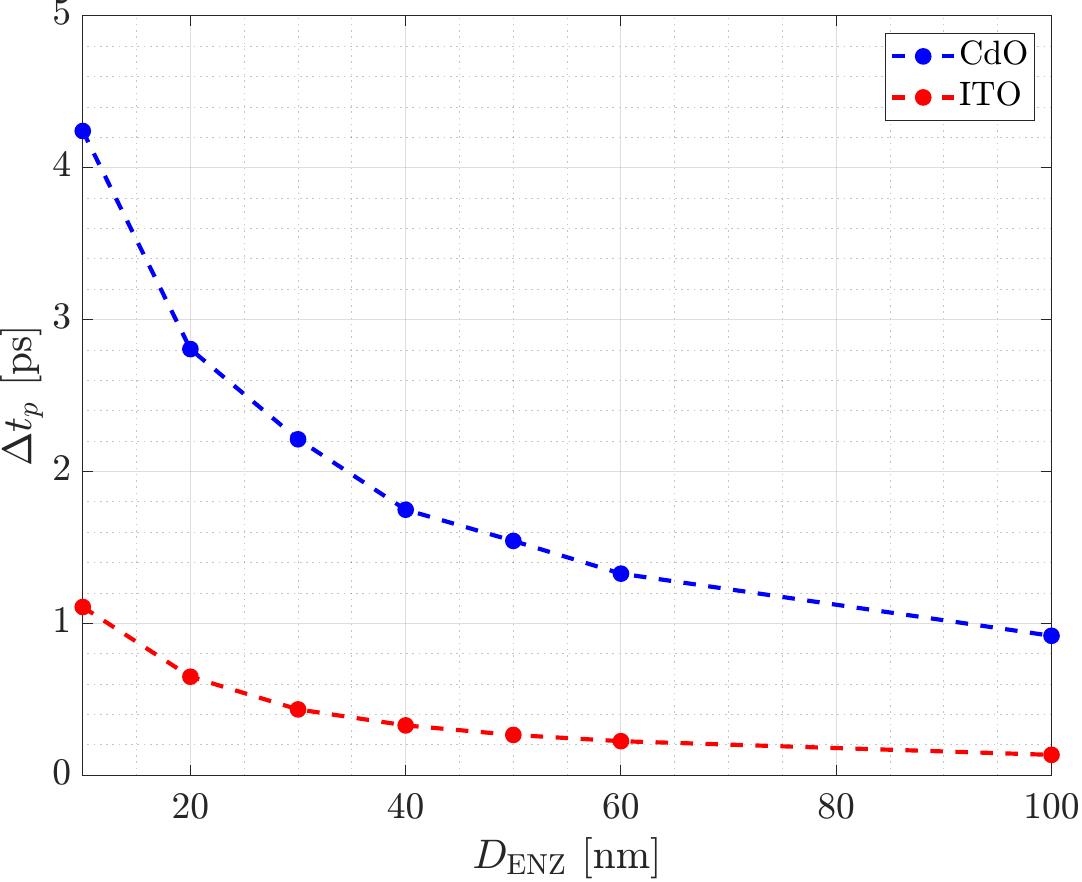}
    \caption{Optimal $\Delta t_p$ as a function of $D_{\rm ENZ}$ for the Si-CdO and SiN-ITO hybrid nanocavities.}
    \label{fig:Dt_p}
\end{figure}

To obtain the pump pulse duration that maximizes $\Delta \omega_{\rm XPM}$, we numerically solve:
\begin{equation}
	\frac{\partial f(\Delta t_p)}{\partial \Delta t_p}=0\,,
\end{equation}
where:
\begin{equation}
    f(\Delta t_p) = c_{\tau} \Delta \omega_p \propto \left(1-e^{-\frac{\Delta t_{p}}{2\tau_p}}\right)^2 \frac{\Delta t_{p}+\tau_p}{\Delta t_{p}+\tau_p +\tau_s} \frac{1}{\Delta t_p}\,.
\end{equation}
It it worth noticing that $f(\Delta t_p)$ is the only part of $\Delta \omega_{\rm XPM}$ that depends on $\Delta t_p$.
In particular, exploiting the fact that $f(\Delta t_p)>0$ we numerically solved $\frac{\partial \ln\left( f(\Delta t_p) \right)}{\partial \Delta t_p}=0$, obtaining the values shown in Fig.~\ref{fig:Dt_p} as a function of $D_{\rm ENZ}$ for the two type of hybrid nanocavities.
It is worth noticing that the optimal pump pulse duration are in the range of picoseconds due to the intrinsic trade-off between the cavity loading dynamics (which favours long pump pulse duration) and the temporal peak intensity (which favours short pump pulse duration).

\section{Material Parameters}

In this section, we summarize the optoelectronic and thermodynamic properties used in our numerical models to characterize the CdO and ITO ENZ nanocavities. Table \ref{tab:material_parameters} reports the specific values adopted for the Drude dispersion model, the third-order nonlinear susceptibilities, and the physical constants driving the thermal and carrier-induced phase noise mechanisms.

\begin{table*}[!hbtp]
    \centering
    \caption{Summary of the physical parameters for CdO and ITO utilized in the theoretical framework and numerical simulations. The dimensionless constants common to both materials, specifically the critical coupling ratio $\kappa_{\text{int}}/\kappa = 0.5$ and the quantum yield of fluctuation $\xi = 1$, are not displayed in the columns.}
    \label{tab:material_parameters}
    \resizebox{\textwidth}{!}{
    \renewcommand{\arraystretch}{1.5}
    \begin{tabular}{lccccccccccccc}
        \hline\hline
        \textbf{Material} & $\chi^{(3)}$ $[\frac{\text{m}^2}{\text{V}^2}]$ & $\varepsilon_b$ & $\gamma$ $[\frac{\text{rad}}{\text{s}}]$ & $\omega_p$ $[\frac{\text{rad}}{\text{s}}]$ & $\rho$ $[\frac{\text{kg}}{\text{m}^3}]$ & $\frac{dn}{dT}$ $[\text{K}^{-1}]$ & $C_v$ $[\frac{\text{J}}{\text{kg}\cdot\text{K}}]$ & $m^*/m_e$ & $\gamma_R$ $[\frac{\text{rad}}{\text{s}}]$ & $\lambda_s$ $[\mu\text{m}]$ & $\lambda_p$ $[\mu\text{m}]$ & $n_s$ & $\chi^{(3)}_{\text{XPM}}$ $[\frac{\text{m}^2}{\text{V}^2}]$ \\
        \hline
        \textbf{CdO} & $7 \times 10^{-16}$ & $5.3$ & $2.80 \times 10^{13}$ & $1.45 \times 10^{15}$ & $8.15 \times 10^3$ & $2 \times 10^{-4}$ & $340$ & $0.15$ & $10^{12}$ & $1.50$ & $3.00$ & $1.990$ & $4.4 \times 10^{-17}$ \\
        \textbf{ITO} & $7.58 \times 10^{-17}$ & $3.8$ & $1.91 \times 10^{14}$ & $3.00 \times 10^{15}$ & $7.00 \times 10^3$ & $1 \times 10^{-4}$ & $380$ & $0.40$ & $10^{12}$ & $0.617$ & $1.234$ & $1.685$ & $4.75 \times 10^{-18}$ \\
        \hline\hline
    \end{tabular}
    }
\end{table*}

\section{Topology-Optimized Nanocavities}

\begin{figure*}[!t]
    \centering
    \includegraphics[width=\textwidth]{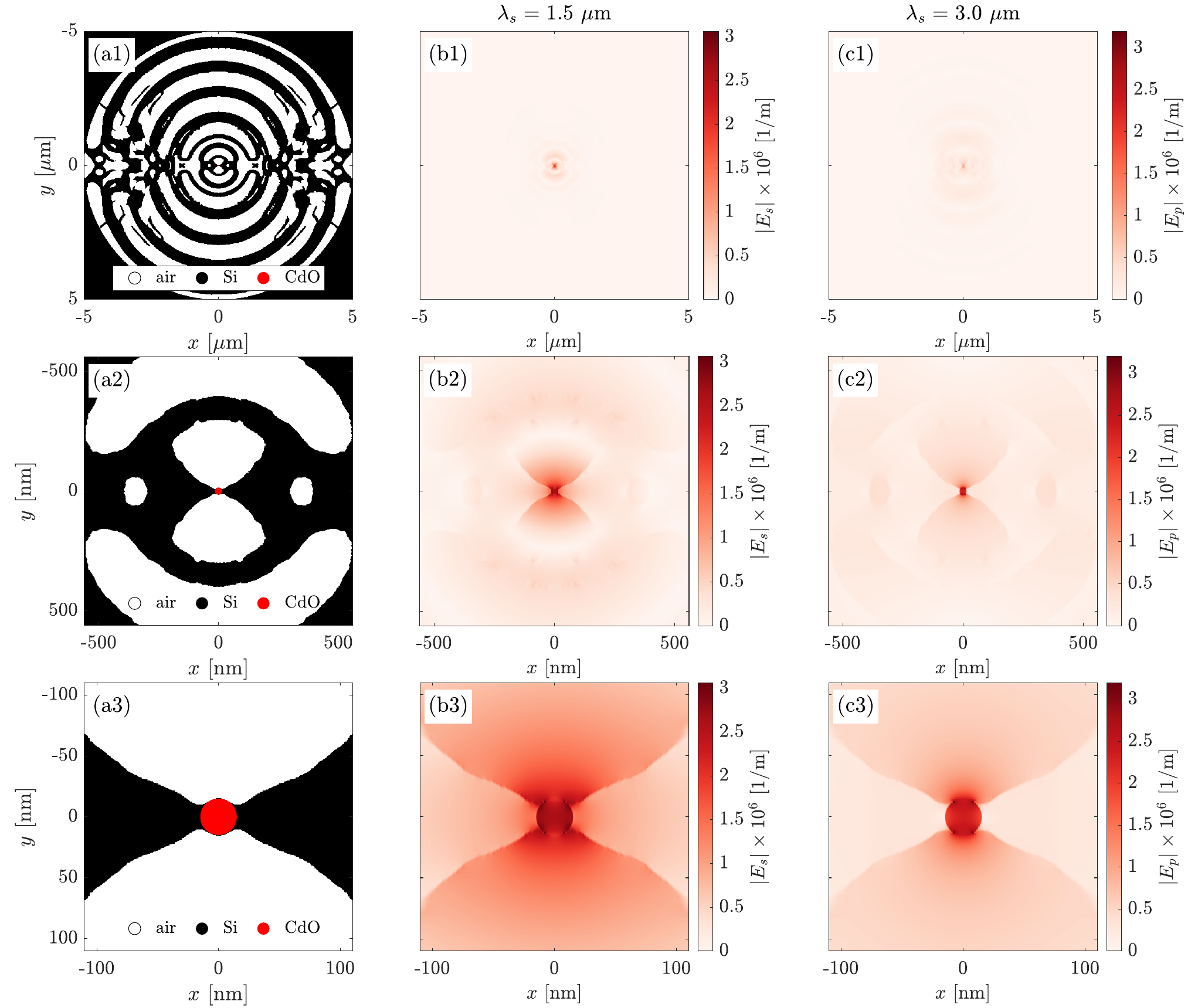}
    \caption{
    Topology-optimized 2D hybrid nanocavity supporting dual-wavelength resonances. (a1–a3) Material distribution of the inverse-designed cavity structure, composed of air (white), silicon (Si, black), and a central cadmium oxide (CdO, red) cylinder with a diameter of $30\,\rm{nm}$. (b1–b3) Spatial distribution of the electric field magnitude ($|\boldsymbol{E}_s|$) for the QNM at $\lambda_s = 1.5\,\mu\text{m}$. (c1–c3) Electric field magnitude ($|\boldsymbol{E}_p|$) for the QNM at $\lambda_p = 3.0\,\mu\text{m}$. The top row (a1–c1) displays the full optimized footprint, while the middle (a2–c2) and bottom (a3–c3) rows provide progressively magnified views of the central region, highlighting the subwavelength field confinement at the core.
    }
    \label{fig:Nanocavity_cdo30}
\end{figure*}

\begin{figure*}[!t]
    \centering
    \includegraphics[width=\textwidth]{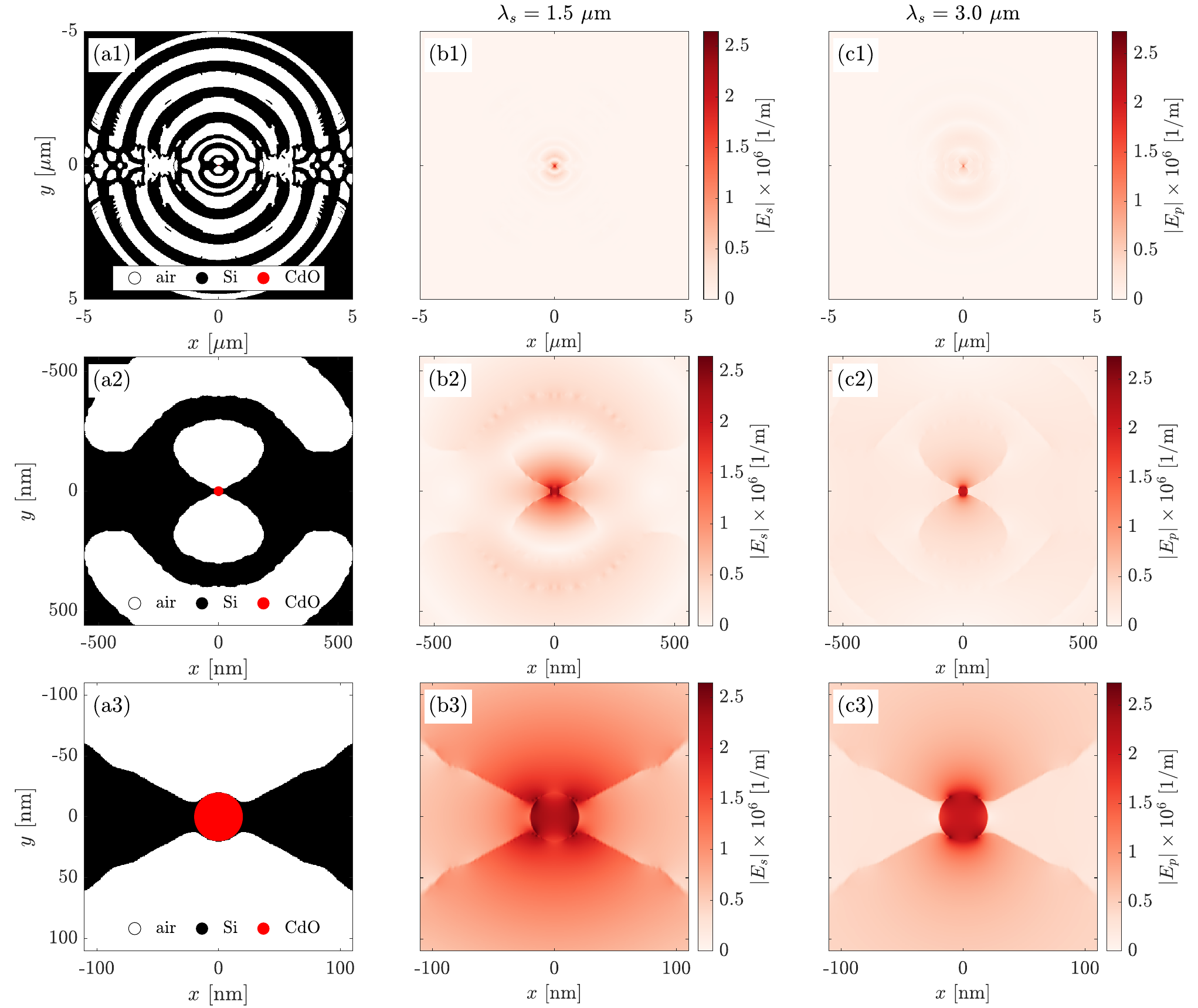}
    \caption{
    Topology-optimized 2D hybrid nanocavity supporting dual-wavelength resonances. (a1–a3) Material distribution of the inverse-designed cavity structure, composed of air (white), silicon (Si, black), and a central cadmium oxide (CdO, red) cylinder with a diameter of $40\,\rm{nm}$. (b1–b3) Spatial distribution of the electric field magnitude ($|\boldsymbol{E}_s|$) for the QNM at $\lambda_s = 1.5\,\mu\text{m}$. (c1–c3) Electric field magnitude ($|\boldsymbol{E}_p|$) for the QNM at $\lambda_p = 3.0\,\mu\text{m}$. The top row (a1–c1) displays the full optimized footprint, while the middle (a2–c2) and bottom (a3–c3) rows provide progressively magnified views of the central region, highlighting the subwavelength field confinement at the core.
    }
    \label{fig:Nanocavity_cdo40}
\end{figure*}

\begin{figure*}[!t]
    \centering
    \includegraphics[width=\textwidth]{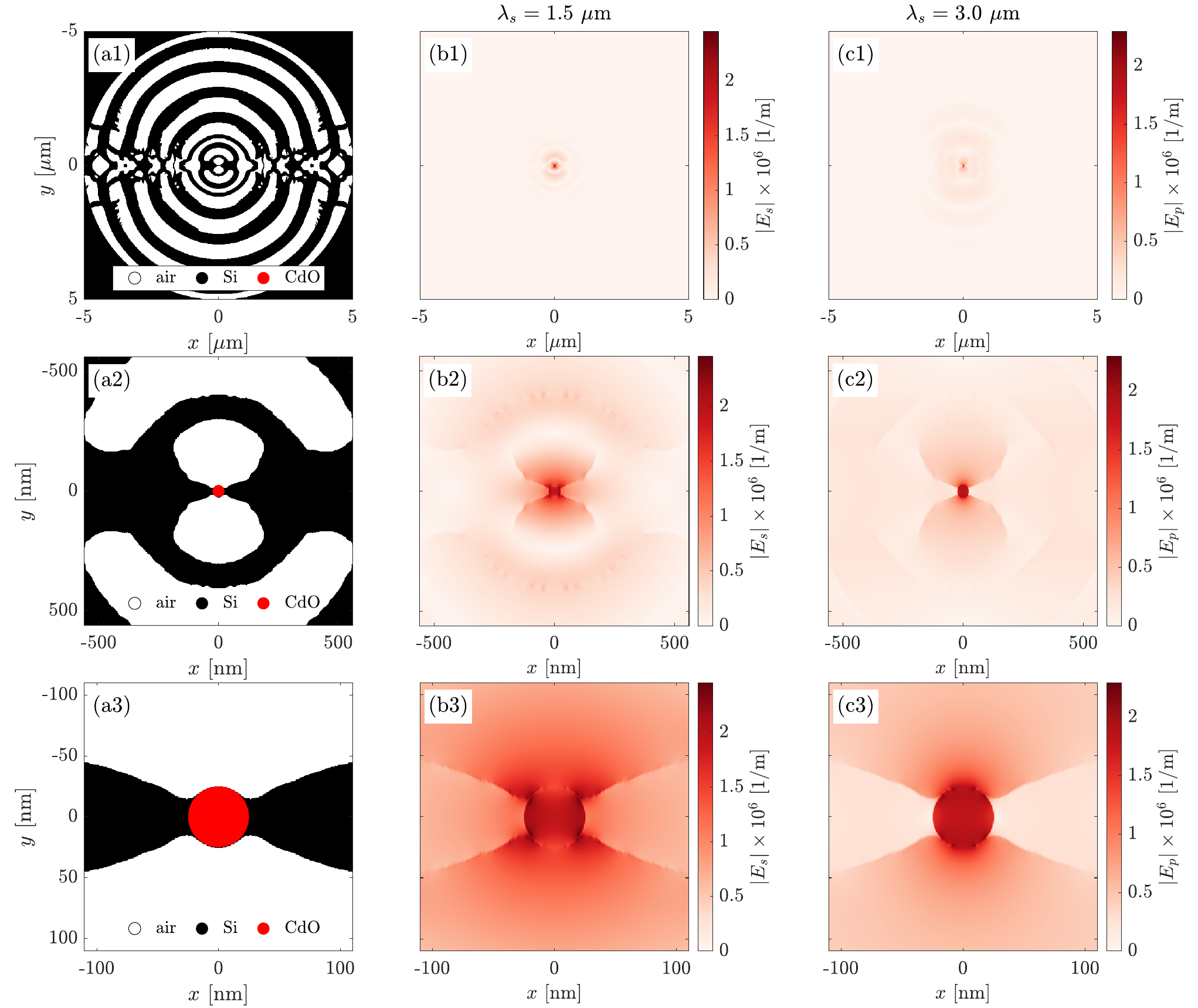}
    \caption{
    Topology-optimized 2D hybrid nanocavity supporting dual-wavelength resonances. (a1–a3) Material distribution of the inverse-designed cavity structure, composed of air (white), silicon (Si, black), and a central cadmium oxide (CdO, red) cylinder with a diameter of $50\,\rm{nm}$. (b1–b3) Spatial distribution of the electric field magnitude ($|\boldsymbol{E}_s|$) for the QNM at $\lambda_s = 1.5\,\mu\text{m}$. (c1–c3) Electric field magnitude ($|\boldsymbol{E}_p|$) for the QNM at $\lambda_p = 3.0\,\mu\text{m}$. The top row (a1–c1) displays the full optimized footprint, while the middle (a2–c2) and bottom (a3–c3) rows provide progressively magnified views of the central region, highlighting the subwavelength field confinement at the core.
    }
    \label{fig:Nanocavity_cdo50}
\end{figure*}

\begin{figure*}[!t]
    \centering
    \includegraphics[width=\textwidth]{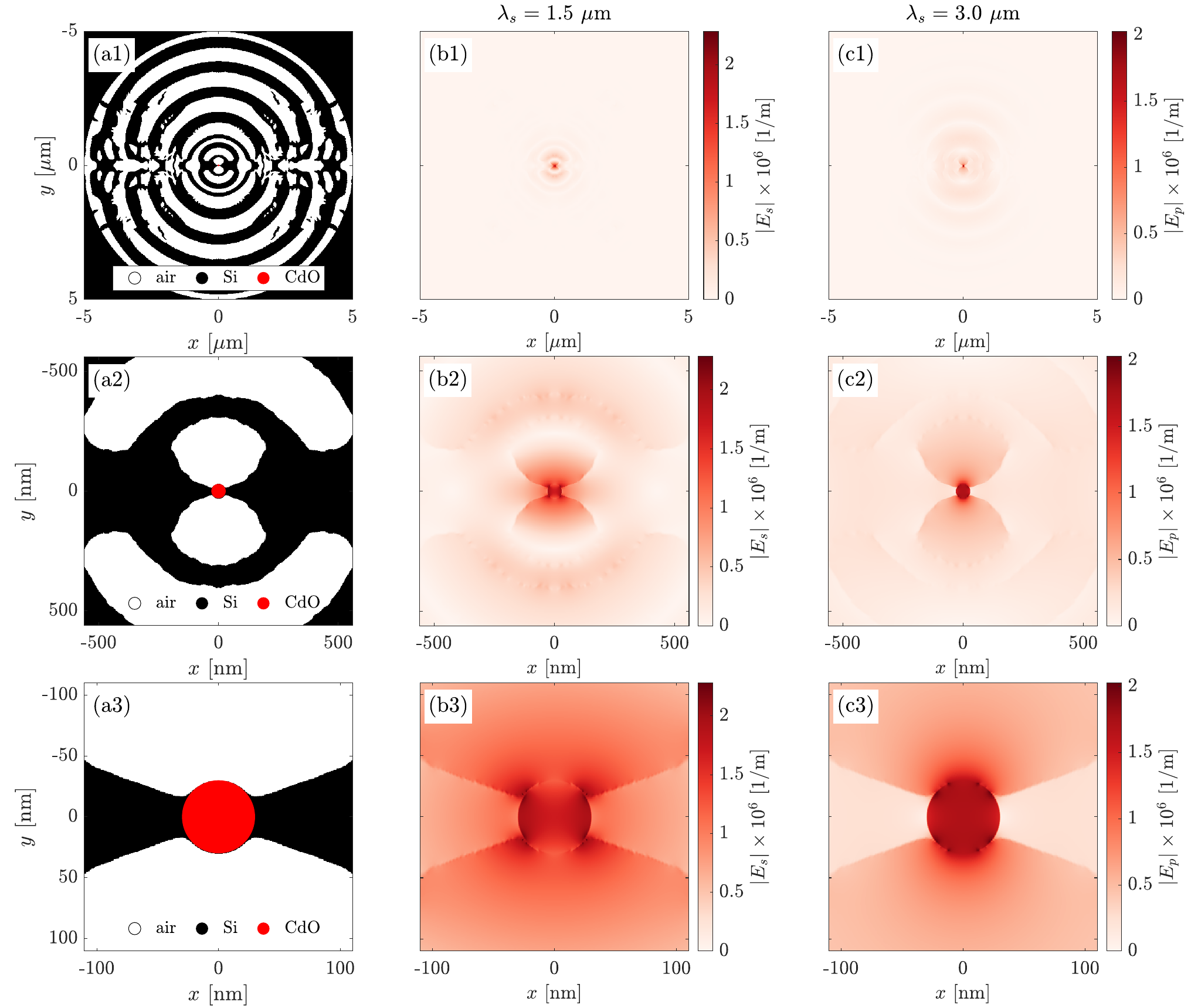}
    \caption{
    Topology-optimized 2D hybrid nanocavity supporting dual-wavelength resonances. (a1–a3) Material distribution of the inverse-designed cavity structure, composed of air (white), silicon (Si, black), and a central cadmium oxide (CdO, red) cylinder with a diameter of $60\,\rm{nm}$. (b1–b3) Spatial distribution of the electric field magnitude ($|\boldsymbol{E}_s|$) for the QNM at $\lambda_s = 1.5\,\mu\text{m}$. (c1–c3) Electric field magnitude ($|\boldsymbol{E}_p|$) for the QNM at $\lambda_p = 3.0\,\mu\text{m}$. The top row (a1–c1) displays the full optimized footprint, while the middle (a2–c2) and bottom (a3–c3) rows provide progressively magnified views of the central region, highlighting the subwavelength field confinement at the core.
    }
    \label{fig:Nanocavity_cdo60}
\end{figure*}

\begin{figure*}[!t]
    \centering
    \includegraphics[width=\textwidth]{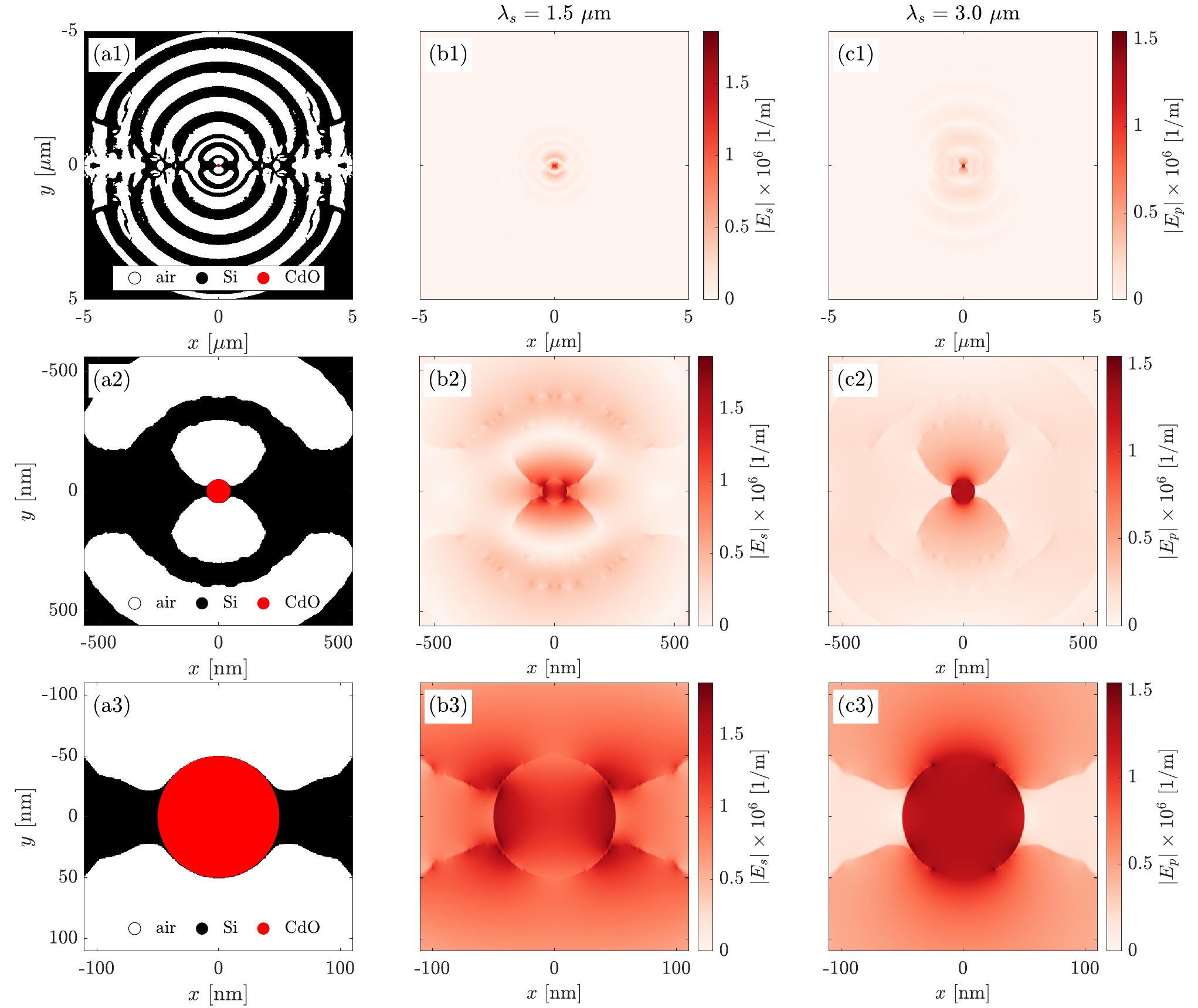}
    \caption{
    Topology-optimized 2D hybrid nanocavity supporting dual-wavelength resonances. (a1–a3) Material distribution of the inverse-designed cavity structure, composed of air (white), silicon (Si, black), and a central cadmium oxide (CdO, red) cylinder with a diameter of $100\,\rm{nm}$. (b1–b3) Spatial distribution of the electric field magnitude ($|\boldsymbol{E}_s|$) for the QNM at $\lambda_s = 1.5\,\mu\text{m}$. (c1–c3) Electric field magnitude ($|\boldsymbol{E}_p|$) for the QNM at $\lambda_p = 3.0\,\mu\text{m}$. The top row (a1–c1) displays the full optimized footprint, while the middle (a2–c2) and bottom (a3–c3) rows provide progressively magnified views of the central region, highlighting the subwavelength field confinement at the core.
    }
    \label{fig:Nanocavity_cdo100}
\end{figure*}

\begin{figure*}[!t]
    \centering
    \includegraphics[width=\textwidth]{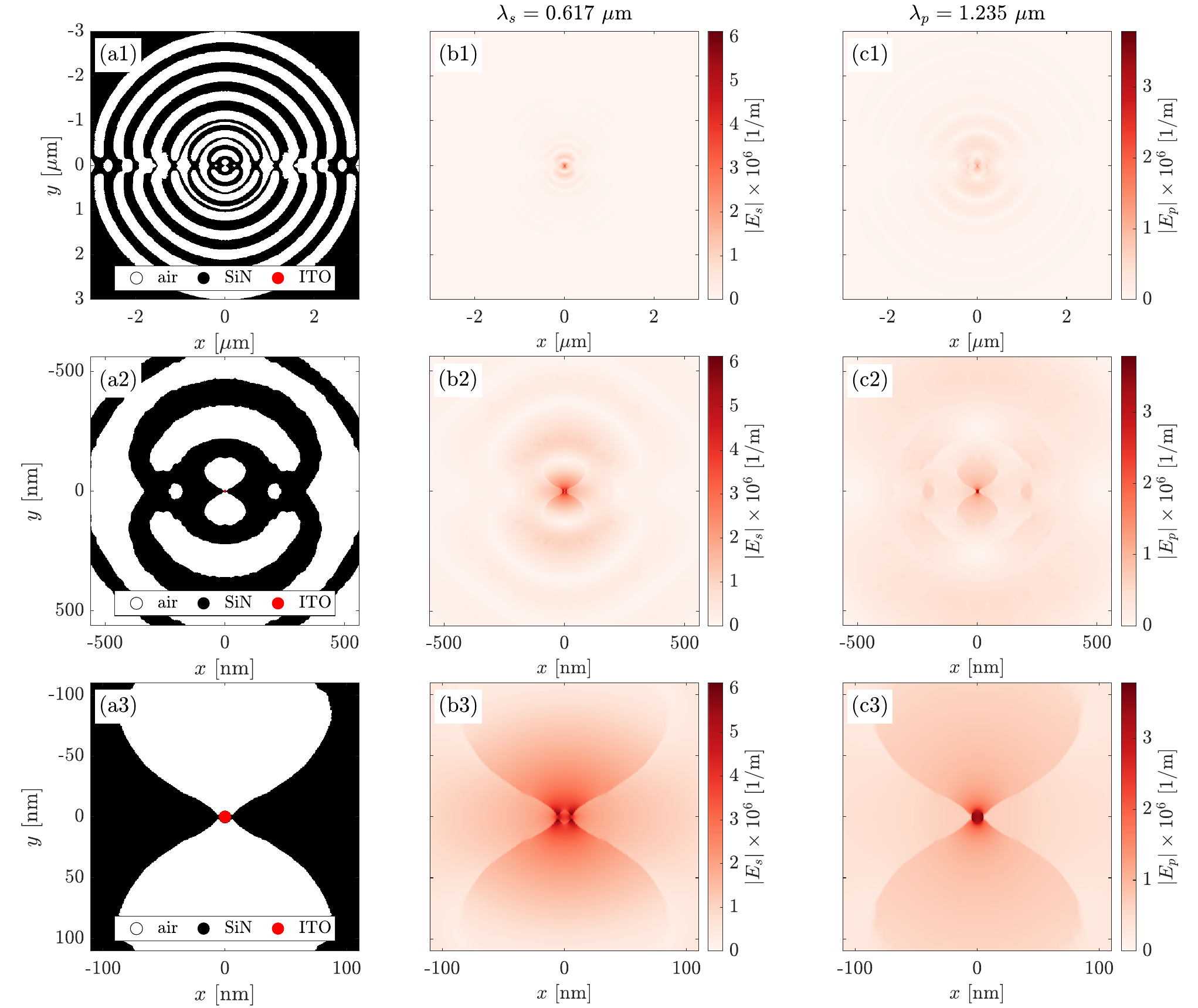}
    \caption{
    Topology-optimized 2D hybrid nanocavity supporting dual-wavelength resonances. (a1–a3) Material distribution of the inverse-designed cavity structure, composed of air (white), silicon nitride (SiN, black), and a central cadmium oxide (ITO, red) cylinder with a diameter of $10\,\rm{nm}$. (b1–b3) Spatial distribution of the electric field magnitude ($|\boldsymbol{E}_s|$) for the QNM at $\lambda_s = 0.617\,\mu\text{m}$. (c1–c3) Electric field magnitude ($|\boldsymbol{E}_p|$) for the QNM at $\lambda_p = 1.235\,\mu\text{m}$. The top row (a1–c1) displays the full optimized footprint, while the middle (a2–c2) and bottom (a3–c3) rows provide progressively magnified views of the central region, highlighting the subwavelength field confinement at the core.
    }
    \label{fig:Nanocavity_ito10}
\end{figure*}

\begin{figure*}[!t]
    \centering
    \includegraphics[width=\textwidth]{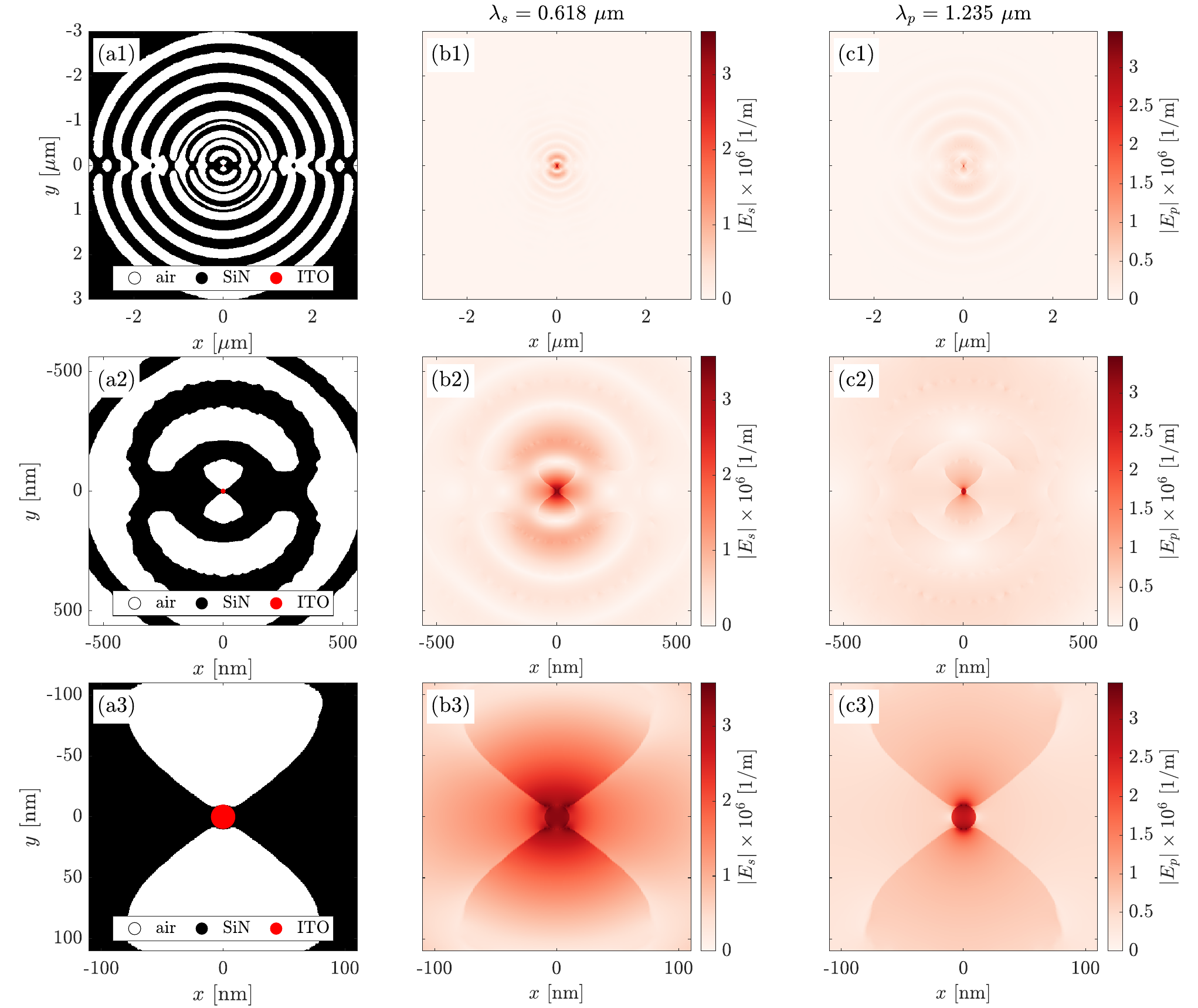}
    \caption{
    Topology-optimized 2D hybrid nanocavity supporting dual-wavelength resonances. (a1–a3) Material distribution of the inverse-designed cavity structure, composed of air (white), silicon nitride (SiN, black), and a central cadmium oxide (ITO, red) cylinder with a diameter of $20\,\rm{nm}$. (b1–b3) Spatial distribution of the electric field magnitude ($|\boldsymbol{E}_s|$) for the QNM at $\lambda_s = 0.618\,\mu\text{m}$. (c1–c3) Electric field magnitude ($|\boldsymbol{E}_p|$) for the QNM at $\lambda_p = 1.235\,\mu\text{m}$. The top row (a1–c1) displays the full optimized footprint, while the middle (a2–c2) and bottom (a3–c3) rows provide progressively magnified views of the central region, highlighting the subwavelength field confinement at the core.
    }
    \label{fig:Nanocavity_ito20}
\end{figure*}

\begin{figure*}[!t]
    \centering
    \includegraphics[width=\textwidth]{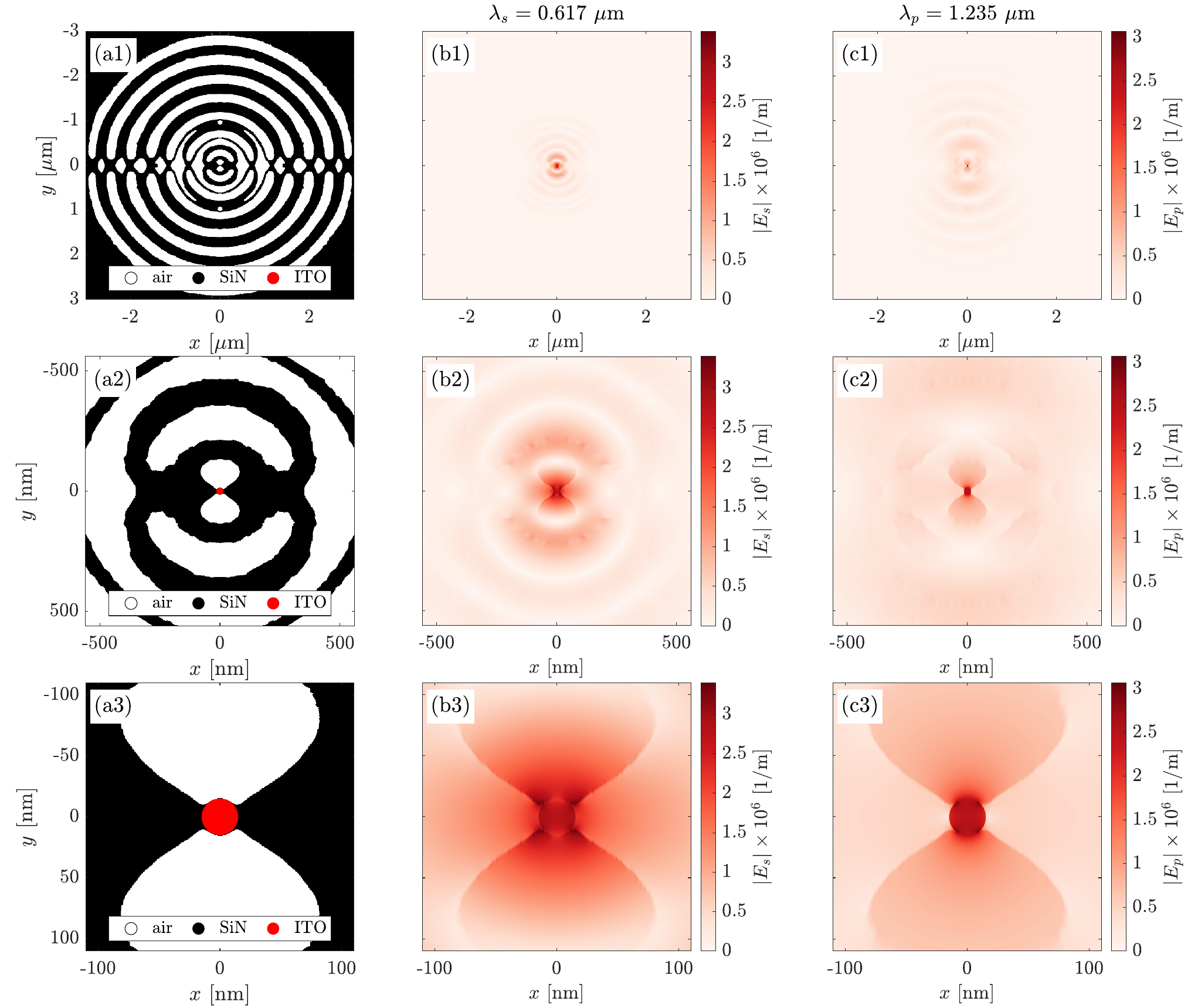}
    \caption{
    Topology-optimized 2D hybrid nanocavity supporting dual-wavelength resonances. (a1–a3) Material distribution of the inverse-designed cavity structure, composed of air (white), silicon nitride (SiN, black), and a central cadmium oxide (ITO, red) cylinder with a diameter of $30\,\rm{nm}$. (b1–b3) Spatial distribution of the electric field magnitude ($|\boldsymbol{E}_s|$) for the QNM at $\lambda_s = 0.617\,\mu\text{m}$. (c1–c3) Electric field magnitude ($|\boldsymbol{E}_p|$) for the QNM at $\lambda_p = 1.235\,\mu\text{m}$. The top row (a1–c1) displays the full optimized footprint, while the middle (a2–c2) and bottom (a3–c3) rows provide progressively magnified views of the central region, highlighting the subwavelength field confinement at the core.
    }
    \label{fig:Nanocavity_ito30}
\end{figure*}

\begin{figure*}[!t]
    \centering
    \includegraphics[width=\textwidth]{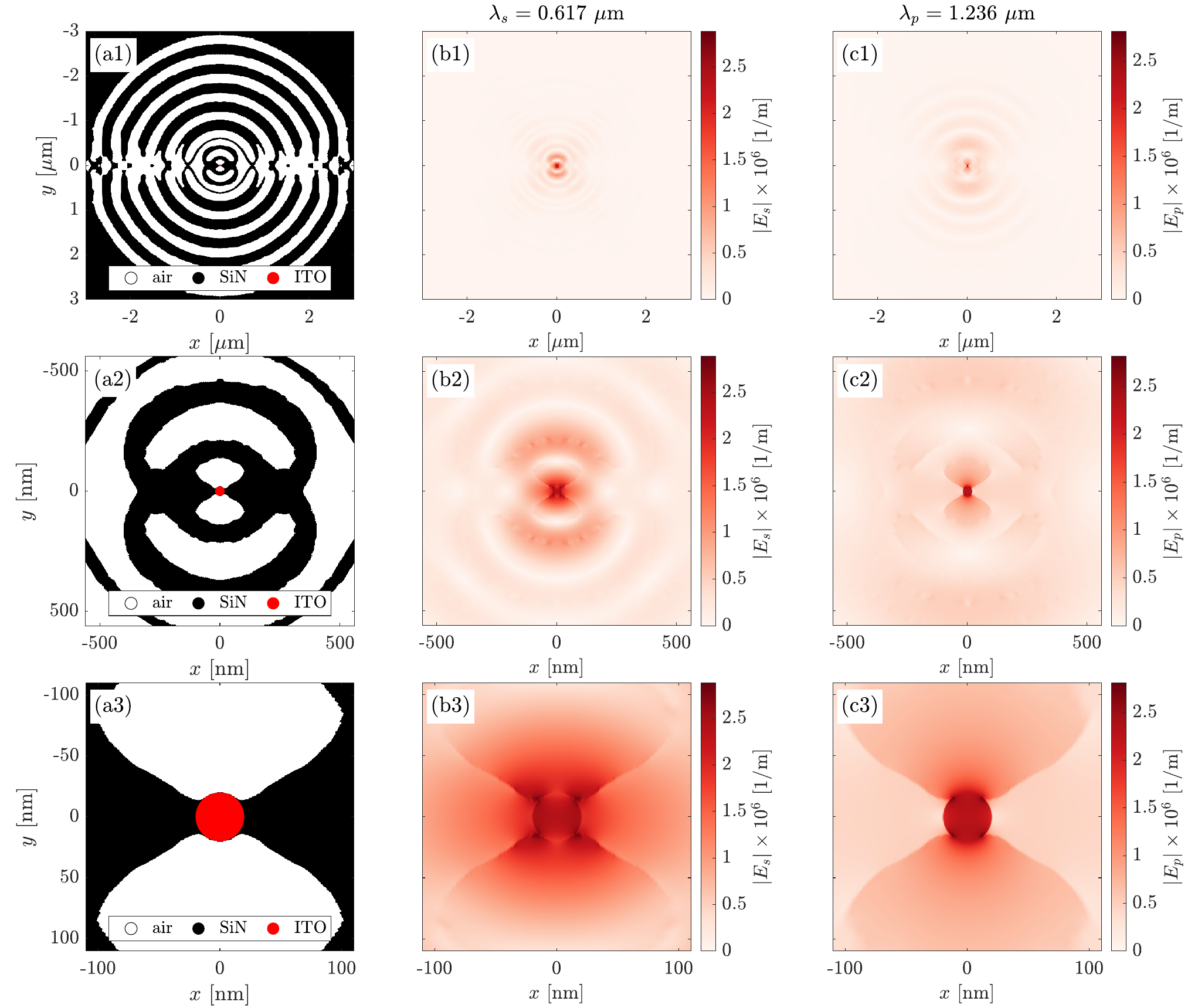}
    \caption{
    Topology-optimized 2D hybrid nanocavity supporting dual-wavelength resonances. (a1–a3) Material distribution of the inverse-designed cavity structure, composed of air (white), silicon nitride (SiN, black), and a central cadmium oxide (ITO, red) cylinder with a diameter of $40\,\rm{nm}$. (b1–b3) Spatial distribution of the electric field magnitude ($|\boldsymbol{E}_s|$) for the QNM at $\lambda_s = 0.617\,\mu\text{m}$. (c1–c3) Electric field magnitude ($|\boldsymbol{E}_p|$) for the QNM at $\lambda_p = 1.236\,\mu\text{m}$. The top row (a1–c1) displays the full optimized footprint, while the middle (a2–c2) and bottom (a3–c3) rows provide progressively magnified views of the central region, highlighting the subwavelength field confinement at the core.
    }
    \label{fig:Nanocavity_ito40}
\end{figure*}

\begin{figure*}[!t]
    \centering
    \includegraphics[width=\textwidth]{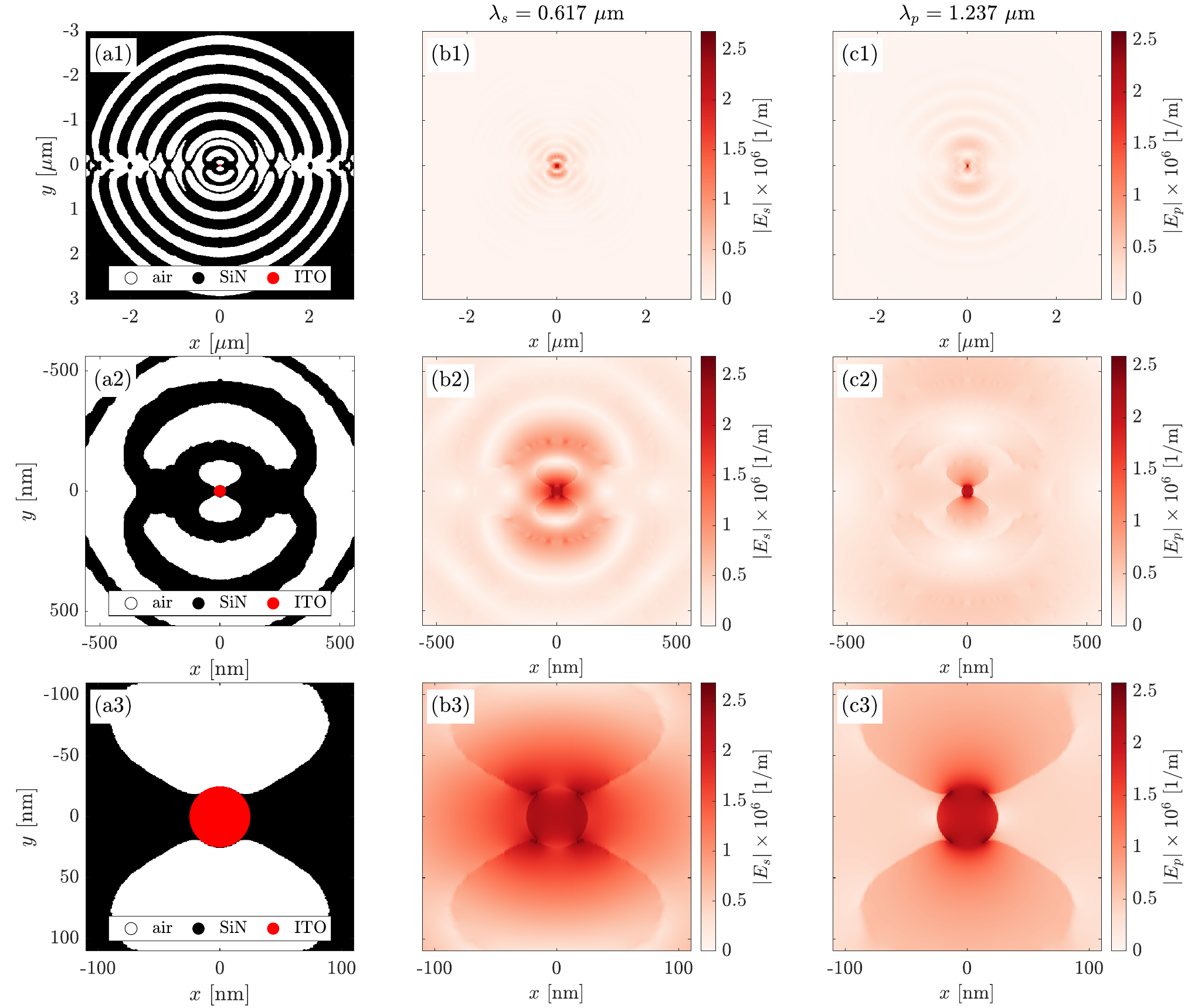}
    \caption{
    Topology-optimized 2D hybrid nanocavity supporting dual-wavelength resonances. (a1–a3) Material distribution of the inverse-designed cavity structure, composed of air (white), silicon nitride (SiN, black), and a central cadmium oxide (ITO, red) cylinder with a diameter of $50\,\rm{nm}$. (b1–b3) Spatial distribution of the electric field magnitude ($|\boldsymbol{E}_s|$) for the QNM at $\lambda_s = 0.617\,\mu\text{m}$. (c1–c3) Electric field magnitude ($|\boldsymbol{E}_p|$) for the QNM at $\lambda_p = 1.237\,\mu\text{m}$. The top row (a1–c1) displays the full optimized footprint, while the middle (a2–c2) and bottom (a3–c3) rows provide progressively magnified views of the central region, highlighting the subwavelength field confinement at the core.
    }
    \label{fig:Nanocavity_ito50}
\end{figure*}

\begin{figure*}[!t]
    \centering
    \includegraphics[width=\textwidth]{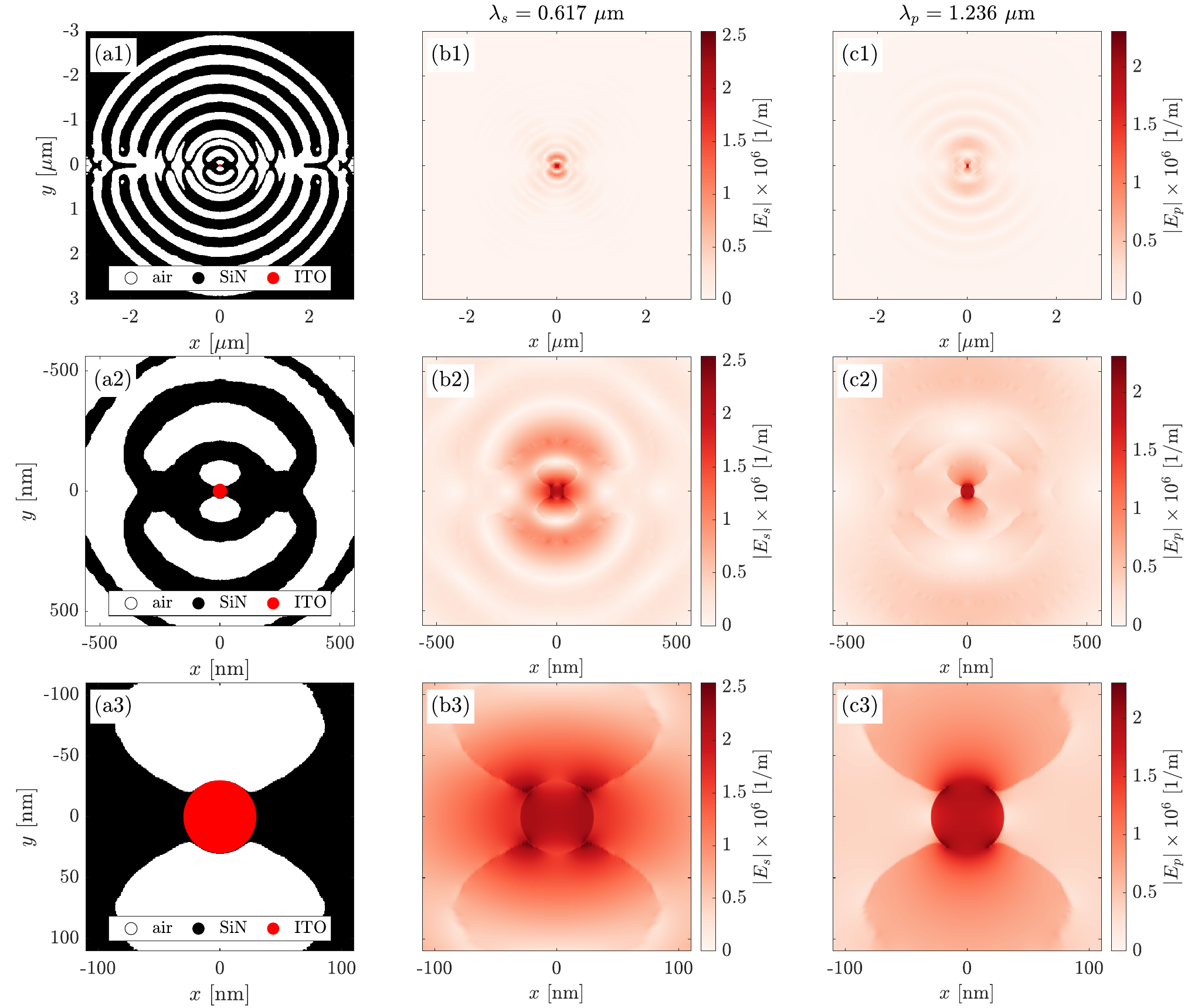}
    \caption{
    Topology-optimized 2D hybrid nanocavity supporting dual-wavelength resonances. (a1–a3) Material distribution of the inverse-designed cavity structure, composed of air (white), silicon nitride (SiN, black), and a central cadmium oxide (ITO, red) cylinder with a diameter of $60\,\rm{nm}$. (b1–b3) Spatial distribution of the electric field magnitude ($|\boldsymbol{E}_s|$) for the QNM at $\lambda_s = 0.617\,\mu\text{m}$. (c1–c3) Electric field magnitude ($|\boldsymbol{E}_p|$) for the QNM at $\lambda_p = 1.236\,\mu\text{m}$. The top row (a1–c1) displays the full optimized footprint, while the middle (a2–c2) and bottom (a3–c3) rows provide progressively magnified views of the central region, highlighting the subwavelength field confinement at the core.
    }
    \label{fig:Nanocavity_ito60}
\end{figure*}

\begin{figure*}[!t]
    \centering
    \includegraphics[width=\textwidth]{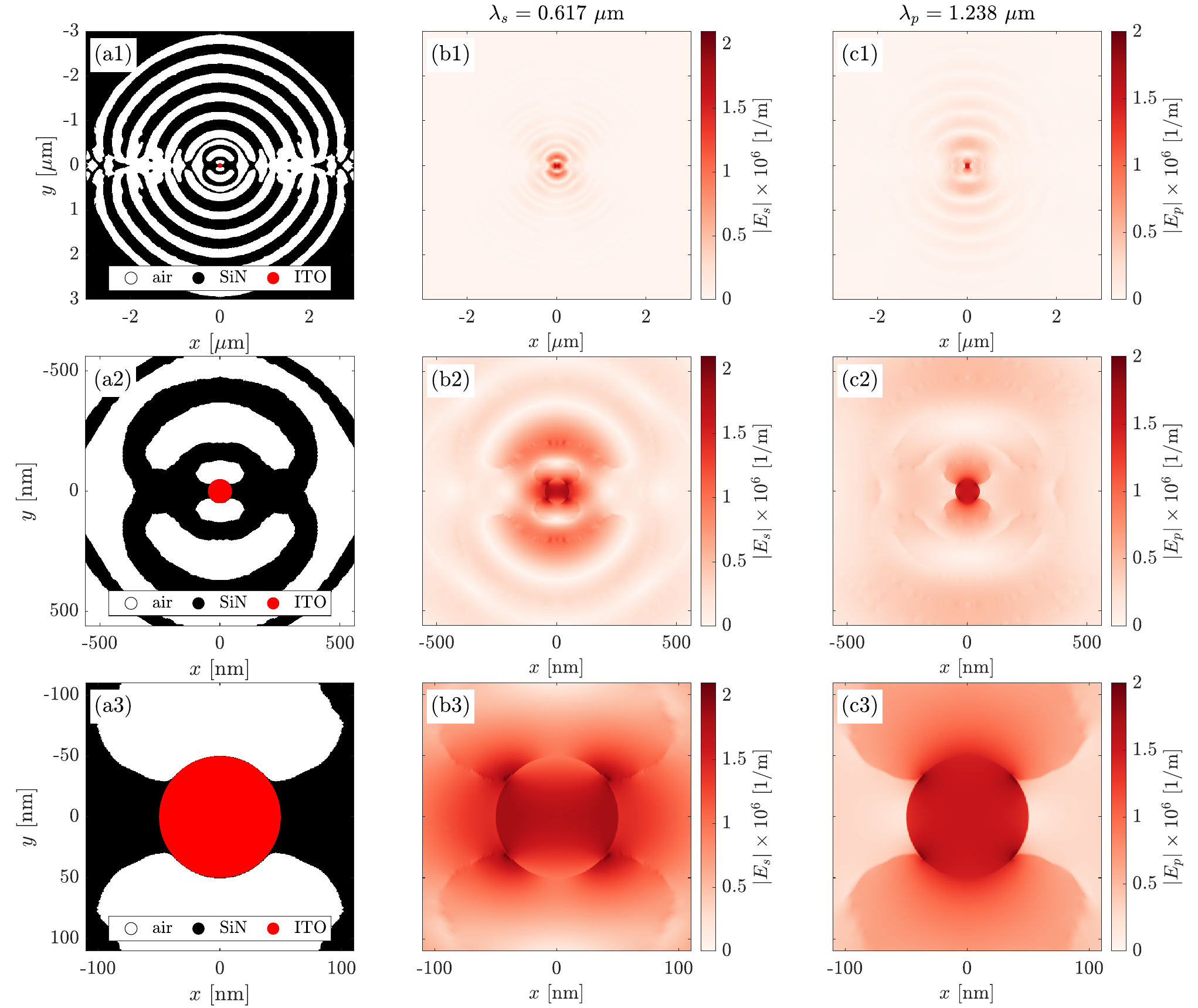}
    \caption{
    Topology-optimized 2D hybrid nanocavity supporting dual-wavelength resonances. (a1–a3) Material distribution of the inverse-designed cavity structure, composed of air (white), silicon nitride (SiN, black), and a central cadmium oxide (ITO, red) cylinder with a diameter of $100\,\rm{nm}$. (b1–b3) Spatial distribution of the electric field magnitude ($|\boldsymbol{E}_s|$) for the QNM at $\lambda_s = 0.617\,\mu\text{m}$. (c1–c3) Electric field magnitude ($|\boldsymbol{E}_p|$) for the QNM at $\lambda_p = 1.238\,\mu\text{m}$. The top row (a1–c1) displays the full optimized footprint, while the middle (a2–c2) and bottom (a3–c3) rows provide progressively magnified views of the central region, highlighting the subwavelength field confinement at the core.
    }
    \label{fig:Nanocavity_ito100}
\end{figure*}

In this section, we provide detailed spatial profiles and material distributions for the topology-optimized dual-wavelength nanocavities discussed in the main text. To demonstrate the robustness and scalability of our free-form inverse design approach, we report numerical results for two distinct hybrid architectures: Si-CdO and SiN-ITO nanocavities. For each material platform, we systematically investigate structural geometries with varying ENZ core diameters $D_{\rm ENZ}$ ranging from $10\,\mathrm{nm}$ to $100\,\mathrm{nm}$.
The following figures (CdO: Figs.~\ref{fig:Nanocavity_cdo10}, \ref{fig:Nanocavity_cdo20}, \ref{fig:Nanocavity_cdo30}, \ref{fig:Nanocavity_cdo40}, \ref{fig:Nanocavity_cdo50}, \ref{fig:Nanocavity_cdo60}, \ref{fig:Nanocavity_cdo100}; ITO: Figs.~\ref{fig:Nanocavity_ito10}, \ref{fig:Nanocavity_ito20}, \ref{fig:Nanocavity_ito30}, \ref{fig:Nanocavity_ito40}, \ref{fig:Nanocavity_ito50}, \ref{fig:Nanocavity_ito60}, \ref{fig:Nanocavity_ito100};) illustrate the extreme sub-wavelength confinement of the electric field achieved simultaneously at the probe and pump wavelengths, which is a fundamental requirement to maximize the XPM efficiency while mitigating radiation losses.

\begin{figure*}[!t]
    \centering
    \includegraphics[width=\textwidth]{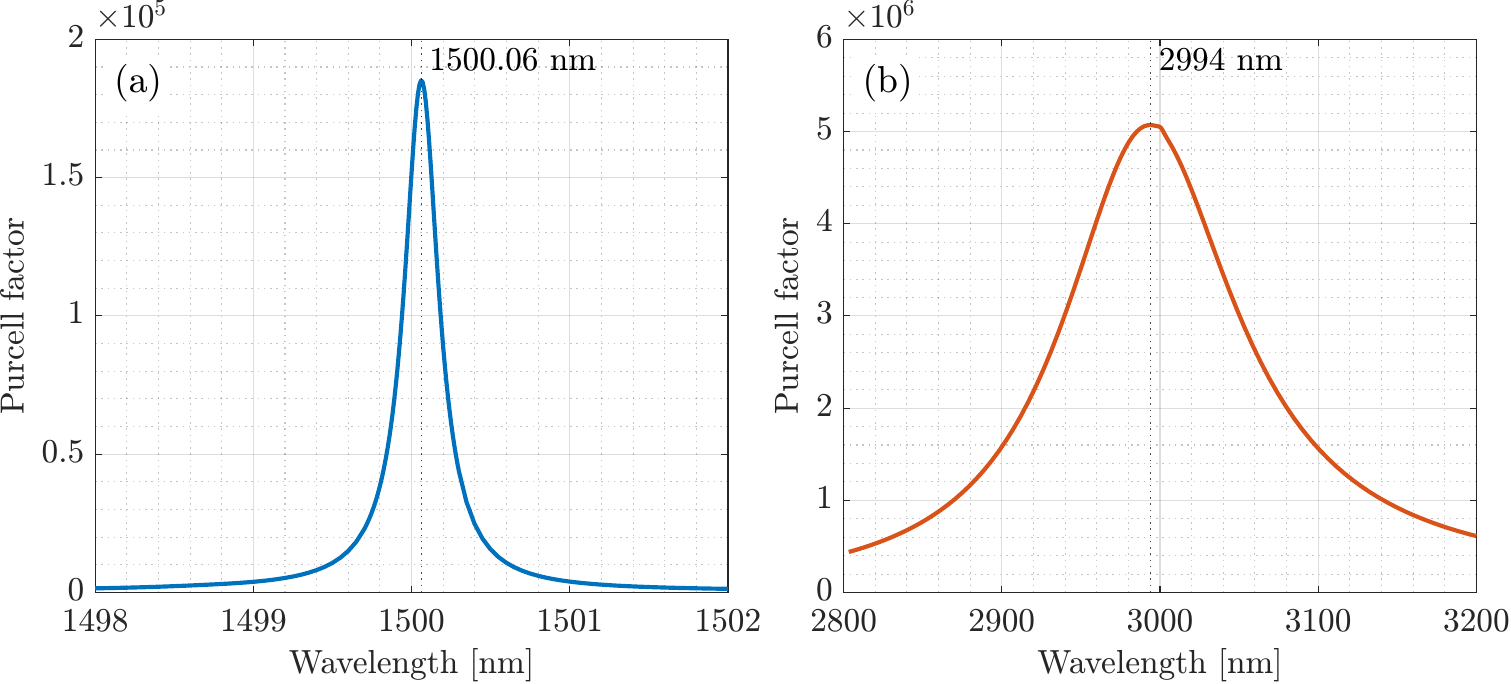}
    \caption{Purcell factor evaluated at the center of the topology-optimized 2D hybrid nanocavity, inside the CdO region of diameter $D_{\rm ENZ}=20\,\text{nm}$. (a) Purcell factor spectrum around $\lambda_s \approx 1.5\,\mu\text{m}$. (b) Purcell factor spectrum around $\lambda_p \approx 3.0\,\mu\text{m}$.}
    \label{fig:Purcell}
\end{figure*}

To further demonstrate the performance of the nanocavity, we evaluated the Purcell factor spectrum at the center of the CdO region for the hybrid Si-CdO nancavity with $D_{\rm ENZ}=20\,{\rm nm}$, as shown in Fig.~\ref{fig:Purcell}.
The distinct resonance peaks around $1.5\ \mu\text{m}$ (Fig.~\ref{fig:Purcell}a) and $3.0\ \mu\text{m}$ (Fig.~\ref{fig:Purcell}b) corroborate the dual-wavelength nature of the device. The large Purcell enhancement observed at both wavelengths is a direct consequence of the successful LDOS maximization, driven by the combination of high quality factors and ultra-small modal volumes achieved through our inverse design strategy.
In Fig.~\ref{fig:Purcell}, the linewidth of the resonance at $\lambda_s \approx 1.5\ \mu\text{m}$ exhibits excellent agreement with the predictions derived from the QNM analysis.
Conversely, the resonance spectrum around $\lambda_p \approx 3.0\ \mu\text{m}$ appears significantly broader than the corresponding QNM quality factor would suggest.
This apparent discrepancy results from the specific excitation conditions used to evaluate the local Purcell factor. In fact, by placing the point dipole exactly at the center of the CdO core---operating within its highly dissipative ENZ regime---the dipole couples strongly not only to the collective hybrid cavity QNM but also to the non-radiative decay channels and highly localized dissipative modes of the CdO material.
Consequently, the local density of optical states broadens due to this near-field quenching effect and the strong material dispersion near the ENZ crossing. Therefore, in any practical realization of the proposed device, engineering strategies to optimally couple the external excitation to the desired hybrid cavity mode, such as far-field radiation pattern optimization \cite{albrechtsen2022nanometer,xiong2024experimental,li2026sub}, will be very important to avoid the damping of these local non-radiative pathways.

\begin{figure*}[!t]
    \centering
    \includegraphics[width=\textwidth]{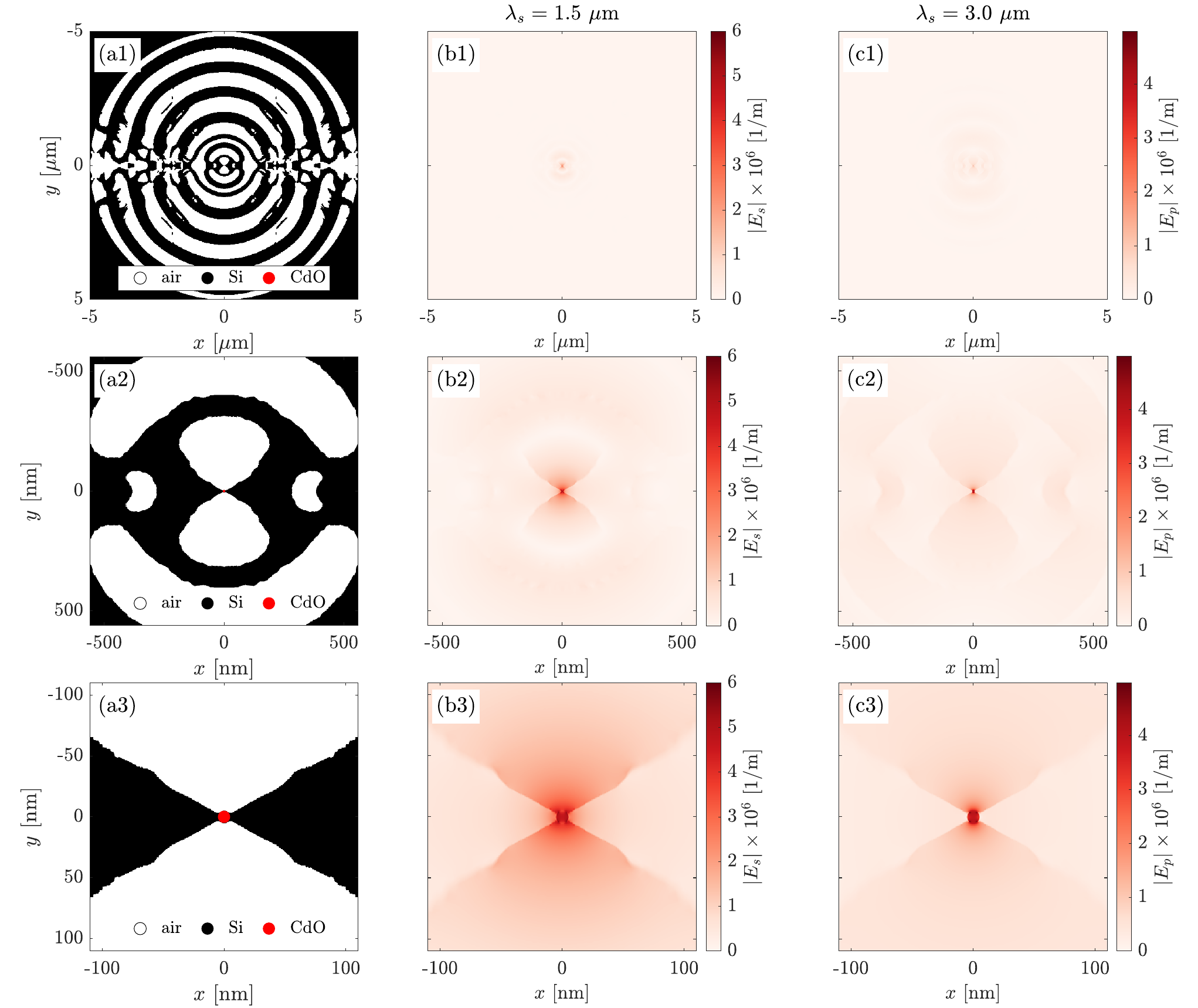}
    \caption{
    Topology-optimized 2D hybrid nanocavity supporting dual-wavelength resonances. (a1–a3) Material distribution of the inverse-designed cavity structure, composed of air (white), silicon (Si, black), and a central cadmium oxide (CdO, red) cylinder with a diameter of $10\,\rm{nm}$. (b1–b3) Spatial distribution of the electric field magnitude ($|\boldsymbol{E}_s|$) for the QNM at $\lambda_s = 1.5\,\mu\text{m}$. (c1–c3) Electric field magnitude ($|\boldsymbol{E}_p|$) for the QNM at $\lambda_p = 3.0\,\mu\text{m}$. The top row (a1–c1) displays the full optimized footprint, while the middle (a2–c2) and bottom (a3–c3) rows provide progressively magnified views of the central region, highlighting the subwavelength field confinement at the core.
    }
    \label{fig:Nanocavity_cdo10}
\end{figure*}

\begin{figure*}[!t]
    \centering
    \includegraphics[width=\textwidth]{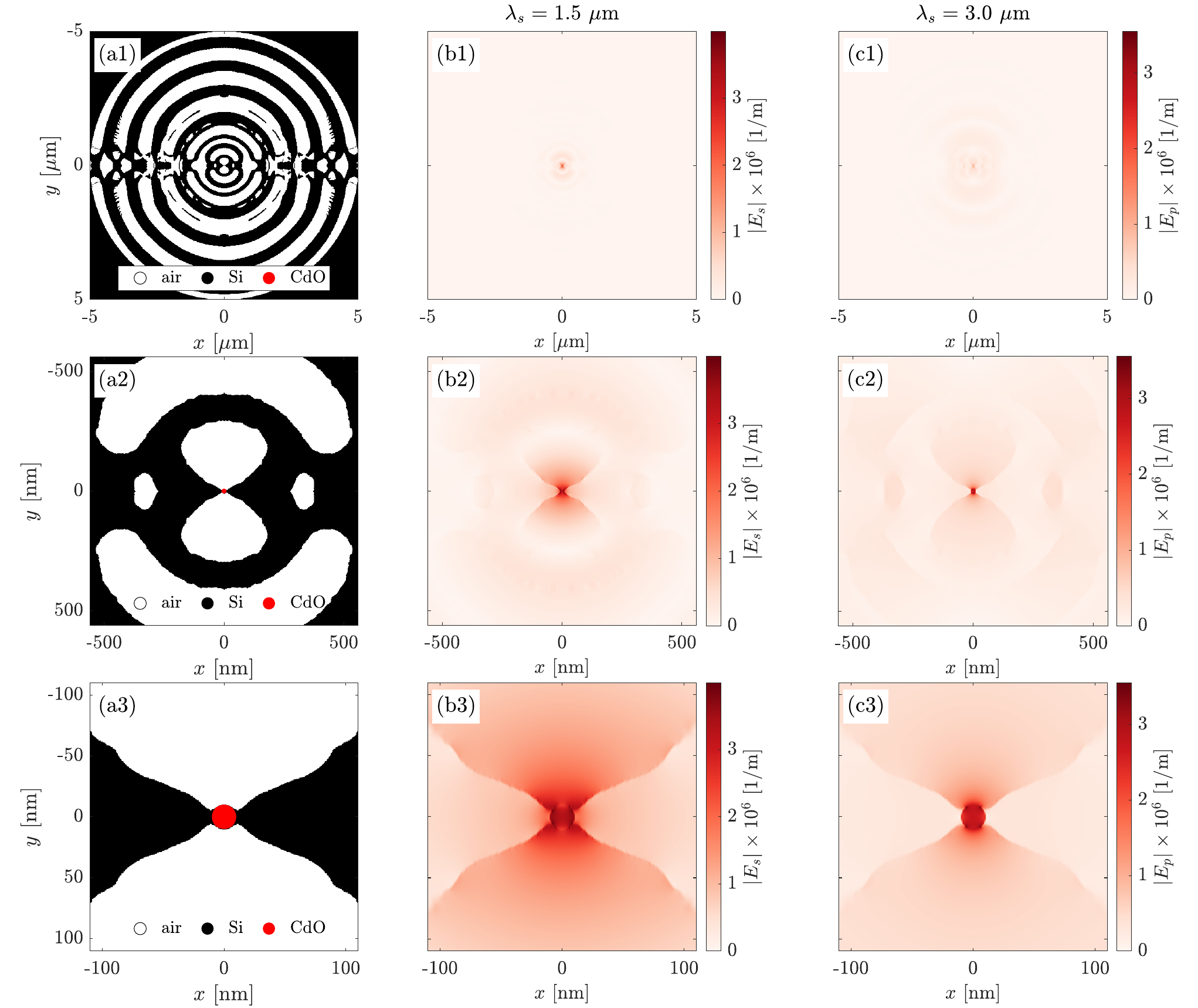}
    \caption{
    Topology-optimized 2D hybrid nanocavity supporting dual-wavelength resonances. (a1–a3) Material distribution of the inverse-designed cavity structure, composed of air (white), silicon (Si, black), and a central cadmium oxide (CdO, red) cylinder with a diameter of $20\,\rm{nm}$. (b1–b3) Spatial distribution of the electric field magnitude ($|\boldsymbol{E}_s|$) for the QNM at $\lambda_s = 1.5\,\mu\text{m}$. (c1–c3) Electric field magnitude ($|\boldsymbol{E}_p|$) for the QNM at $\lambda_p = 3.0\,\mu\text{m}$. The top row (a1–c1) displays the full optimized footprint, while the middle (a2–c2) and bottom (a3–c3) rows provide progressively magnified views of the central region, highlighting the subwavelength field confinement at the core.
    }
    \label{fig:Nanocavity_cdo20}
\end{figure*}

%\bibliography{references}

% Stampa la bibliografia anche per questo documento
%\putbib[references]

%\end{bibunit} % <-- Fine della seconda unità

\end{document}